\documentclass[journal]{IEEEtran}

\ifCLASSINFOpdf
   \usepackage[pdftex]{graphicx}
\else
   \usepackage[dvips]{graphicx}
\fi

\usepackage{mathtools}
\usepackage{url}
\usepackage{amsmath}
\usepackage{amsfonts}
\usepackage{amssymb}
\usepackage{algorithm}
\usepackage{algorithmic}
\usepackage{cite}
\usepackage{graphicx}
\usepackage{subfigure}
\usepackage{color}
\usepackage{bm}
\usepackage{multirow}

\begin{document}
%
\title{Energy-Efficient Resource Allocation for Secure NOMA-Enabled Mobile Edge Computing Networks}
%
%
%

\author{Wei~Wu, \IEEEmembership{Member,~IEEE,}
        Fuhui Zhou, \IEEEmembership{Member,~IEEE,}
        Rose Qingyang Hu, \IEEEmembership{Senior Member, IEEE,} and
        Baoyun Wang, \IEEEmembership{Member,~IEEE}
\thanks{This paper was presented in part at the IEEE International Conference on Communications (ICC), Shanghai, China, May
2019 \cite{IEEEhowto:35}.}
\thanks{W. Wu is with the College of Communication and Information Engineering, Nanjing University of Posts and Telecommunications,
Nanjing, 210003, China (e-mail: weiwu@njupt.edu.cn).}
\thanks{F. Zhou is with the Department of Electrical and Computer Engineering as a Research Follow at Utah State University, U.S.A. F. Zhou is also with the School of Information Engineering, Nanchang University, 330031, China (e-mail: zhoufuhui@ieee.org)}
\thanks{R. Q. Hu is with the Department of Electrical and Computer Engineering, Utah State University, USA. (email: rose.hu@usu.edu).}
\thanks{B. Wang is with the College of Overseas Education, Nanjing University of Posts and Telecommunications,
Nanjing, 210023, China (e-mail: bywang@njupt.edu.cn). \textit{(Corresponding author: Fuhui Zhou.)}}}

\maketitle

\begin{abstract}
Mobile edge computing (MEC) has been envisaged as a promising technique in the next-generation wireless networks. In order to improve the security of computation tasks offloading and enhance user connectivity, physical layer security and non-orthogonal multiple access (NOMA) are studied in MEC-aware networks. The secrecy outage probability is adopted to measure the secrecy performance of computation offloading by considering a practically passive eavesdropping scenario. The weighted sum-energy consumption minimization problem is firstly investigated subject to the secrecy offloading rate constraints, the computation latency constraints and the secrecy outage probability constraints. The semi-closed form expression for the optimal solution is derived. We then investigate the secrecy outage probability minimization problem by taking the priority of two users into account, and characterize the optimal secrecy offloading rates and power allocations with closed-form expressions. Numerical results demonstrate that the performance of our proposed design are better than those of the alternative benchmark schemes.
\end{abstract}

\begin{IEEEkeywords}
Mobile edge computing, non-orthogonal multiple access, physical layer security, secrecy outage probability, partial offloading.
\end{IEEEkeywords}

%
\IEEEpeerreviewmaketitle

\section{Introduction}

\IEEEPARstart{T}{he} rapid development of next-generation wireless networks has spawned the unprecedented proliferation of smart devices (e.g., tablet computers, smart phones, smart furniture and wearable devices) and new applications (e.g., augmented reality, autonomous driving, and tele-surgery) \cite{IEEEhowto:35, IEEEhowto:30}. With the massive deployment of smart devices, how to accommodate them with limit resources is a challenging task \cite{IEEEhowto:31}. Moreover, a lot of new emerging applications can be highly computation-intensive and latency-sensitive, making it a very challenging task for the power limited and size constrained terminal devices to deliver the desirable quality of service in these circumstances.

In order to tackle the above-mentioned challenges, mobile edge computing (MEC) and non-orthogonal multiple access (NOMA) have been envisaged as two promising techniques in the next-generation wireless networks \cite{IEEEhowto:1, IEEEhowto:2, IEEEhowto:3}. In an MEC system, distributed MEC servers are dedicatedly deployed in a close proximity to the terminal devices that can offload partial or all of their computation tasks to the MEC servers for computing. Therefore, MEC enables the cloud-like computing for the small-size and low-power terminal devices in a cost-effective and low-latency manner \cite{IEEEhowto:3, IEEEhowto:32, IEEEhowto:4, IEEEhowto:5}. As a potential key technology in the fifth generation (5G) networks, NOMA brings fundamental changes to the regime of multiple access and achieves a much higher spectral efficiency than the conventional orthogonal multiple access (OMA) by implementing advanced transceiver designs, such as superposition coding and successive interference cancellation (SIC) \cite{IEEEhowto:6, IEEEhowto:7}.

Recently, MEC has attracted ever-increasing research interests in both industry and academia due to its powerful capability in facilitating the real-time implementation of computation-extensive tasks. To fully reap the advantage of MEC, the joint design of communication and computation resource allocation is a critical issue that should be properly addressed \cite{IEEEhowto:8, IEEEhowto:9, IEEEhowto:10, IEEEhowto:11}. For example, the authors in \cite{IEEEhowto:8} proposed a novel joint communication and computation cooperation approach by introducing an additional helper acting as the auxiliary computing server and the decode-and-forward (DF) relay. However, the finite battery lifetime of the size-constrained end devices causes longstanding performance limitations of the MEC networks. To resolve this issue, recent literatures \cite{IEEEhowto:9, IEEEhowto:10, IEEEhowto:11} studied the integration of wireless power transfer (WPT) into MEC networks, and envisioned significant computation performance improvement for both  \textit{partial} \cite{IEEEhowto:9} or \textit{binary} \cite{IEEEhowto:10, IEEEhowto:11} offloading modes. Moreover, in \cite{IEEEhowto:10}, a more challenging multi-user MEC scenario was considered, where the multi-user computing mode selection and strong coupling with transmission time allocation problems were tackled by the alternating direction method of multipliers decomposition technique. While in \cite{IEEEhowto:11}, the unmanned aerial vehicle (UAV)-enabled MEC was considered due to the existence of severe propagation loss of terrestrial communications.

Realizing the superiority of NOMA in spectrum utilization, the application of NOMA to MEC has recently received extensive attention \cite{IEEEhowto:2, IEEEhowto:4, IEEEhowto:13, IEEEhowto:14, IEEEhowto:36, IEEEhowto:37}. Wang \textit{et al}. \cite{IEEEhowto:4} first investigated the application of NOMA uplink transmission to MEC. A joint SIC decoding order, communication and computing resource allocation scheme was proposed in multi-user MEC networks. It was shown that the proposed scheme can achieve a higher energy efficiency than the OMA-based and other benchmark offloading schemes. To support the massive connectivity requirement of 5G wireless networks, a novel NOMA augmented edge computing model was considered \cite{IEEEhowto:13}, where the user clustering, frequency and computing resource allocation were jointly designed with traditional decision variables. Different from these prior works that studied NOMA-assisted MEC via optimization frameworks, Ding \textit{et al}. \cite{IEEEhowto:2} presented a comprehensive theoretic performance analysis of the impact of both NOMA uplink transmission and downlink transmission on MEC. Diverse asymptotic studies revealed the unique role of the users' channel conditions and transmit powers on the application of NOMA to MEC. Subsequently, Ding \textit{et al}. \cite{IEEEhowto:14} further studied the energy consumption of NOMA-assisted MEC offloading by jointly optimizing the power and time allocation. Based on the obtained closed-form expressions, Ding \textit{et al}. \cite{IEEEhowto:14} revealed the important properties of NOMA-MEC offloading by comparing the performance among hybrid-NOMA-MEC, pure NOMA-MEC and OMA-MEC under different task delay tolerances. Considering the limited computation capability of the MEC server, Zeng \textit{et al}. \cite{IEEEhowto:36} investigated the joint design of subcarrier, transmission power and computational resource allocation to minimize the energy consumption at the users. Furthermore, the overall delay, which includes the mobile terminal's local computation delay, the round trip delay and the edge server's computation delay, minimization problem was studied by Wu \textit{et al}. in \cite{IEEEhowto:37}. Through exploiting the layered structure of the delay minimization problem, multiple algorithms were proposed to obtain the optimal offloading solution jointly.

On the other hand, owing to the broadcast nature of wireless communication, the task offloading from end devices to the access point (AP) over wireless channels is vulnerable to malicious attacks that result in information leakage. Therefore, it is crucial to take the security issue into account for the success of MEC. Physical layer security has been widely envisioned to be an effective wireless information security transmission protection technique \cite{IEEEhowto:15}. The perfect secure data transmission can be guaranteed once the channel state information (CSI) of the wiretap channel is available at the legitimate users (see, e.g., \cite{IEEEhowto:16, IEEEhowto:17}), and the robust security is absolutely achievable despite the imperfect CSI (see, e.g. \cite{IEEEhowto:18, IEEEhowto:19}). Based on this, Xu \textit{et al}. \cite{IEEEhowto:20} first proposed to employ the physical layer security to secure the MEC offloading in the practical imperfect CSI scenario, where the multiuser subcarrier allocation problem was studied and new secure issues were introduced by keeping the offloading rate at each user not exceed its secrecy rate to the AP.

From the above discussion, we note that the design problem of NOMA-assisted MEC against external eavesdropper with appropriate secrecy and quality of service (QoS) performance metrics has yet been investigated. Moreover, the security issue is of crucial importance to the success of MEC. And, the perfect knowledge of external eavesdropper's channel state information
is practically unknown to the AP. These factors motivate us to design NOMA-assisted secure offloading schemes for the practical scenario where the transmitter does not know the eavesdropper's instantaneous channel state information. It can help us further to expand the application of NOMA and gain better understanding of MEC offloading security.

In this paper, we consider an uplink NOMA-based MEC system consisting of one AP integrated with an MEC server, multiple end users and an external eavesdropper. Under the NOMA and partial offloading setup, all the users can simultaneously offload partial computation tasks to the AP over the same resource (time/frequency) block. Since the passive eavesdropper's instantaneous CSI cannot be known by the AP in practice, we take the secrecy outage probability as the secrecy metric to measure the secrecy offloading performance of the NOMA-based MEC network. Please note that in \cite{IEEEhowto:35}, we only studied the weighted sum-energy consumption minimization problem under the secrecy offloading rate constraint of each user. Moreover, we did not study the secrecy outage probability minimization problems with given latency and energy budget \cite{IEEEhowto:35}. The potential applications of our considered secure NOMA enabled MEC system can be the MEC-aware NOMA narrowband Internet of Things (IoT) networks with densely deployed access points and mobile terminals \cite{IEEEhowto:3}. The access points in the networks are responsible for dual functions of information transmission and edge computing service while the mobile terminals have the function of information transmission/reception.The primary contributions of this paper are summarized as follows:

\begin{itemize}
  \item	We comprehensively investigate the design of NOMA-based MEC networks against the external eavesdropper. An innovative design framework is developed by jointly optimizing the number of locally computed bits, the power allocation, the codeword transmission rates and the confidential data rates at the uplink users. Note that the number of offloaded bits of each user is characterized by the confidential data rate of each user, which appropriately captures the real rate of informative data received at the AP.
  \item	Targeting at an energy-efficient secure NOMA-MEC design, we minimize the users' weighted sum-energy consumption subject to the secrecy offloading rate constraints, the computation latency constraints and the secrecy outage probability constraints. The problem is challenge non-convex. And unlike many traditional design problems of MEC, our problem cannot be transformed into a sequence of linear programs (LPs) or find its Lagrange dual problem with strong duality due to the secrecy outage probability constraint and the coupling among optimization variables, which make it more difficult to solve. Leveraging the state-of-the-art optimization approaches, we obtain the optimal solution in a semi-closed form.
  \item	We further focus on the problem of minimizing the secrecy outage probability by taking the priority of multiple users into account, which has never been investigated in the literature. Through analysis and transformation, we derived the optimal secrecy offloading rates and power allocations in closed-form expressions. We find that the secure offloading outage event of the secondary priority user occurs constantly when the first priority user's transmission power is large enough. Moreover, we also characterize how channel gain and transmission power influence the secrecy outage performance.
  \item	Extensive numerical results are provided to evaluate the performance of our proposed design. We compare the performance of the secure NOMA-MEC scheme with that of the secure NOMA full offloading scheme and the secure OMA-MEC scheme. The conventional design without an eavesdropper is also introduced as a performance upper bound. It is shown that our proposed design can significantly reduce the energy consumption and the secrecy outage probability compared with two benchmark schemes.
\end{itemize}

The rest of this paper is organized as follows. Section II describes the system model. Section III focuses on the weighted sum-energy consumption minimization problem subjected to secrecy offloading considerations. Section IV minimizes the secrecy outage probability of the uplink users based on preset priority. Numerical results are provided in Section V. Finally, our paper is concluded in Section VI.

\emph{Notations:} Vectors are represented by boldface letters. $\mathbb{E}\left\{\cdot\right\}$ denotes the statistical expectation. $|\cdot|$ represents the absolute value of a complex scalar. $x \sim \mathcal{CN}\left( {a,\;b} \right)$ means that the scalar $x$ follows a complex Gaussian distribution with mean $a$ and covariance $b$.

\section{System Model}

\begin{figure}[!t]
  {
  \centering
  \centerline{\includegraphics[width=3.6in,height=2.4in]{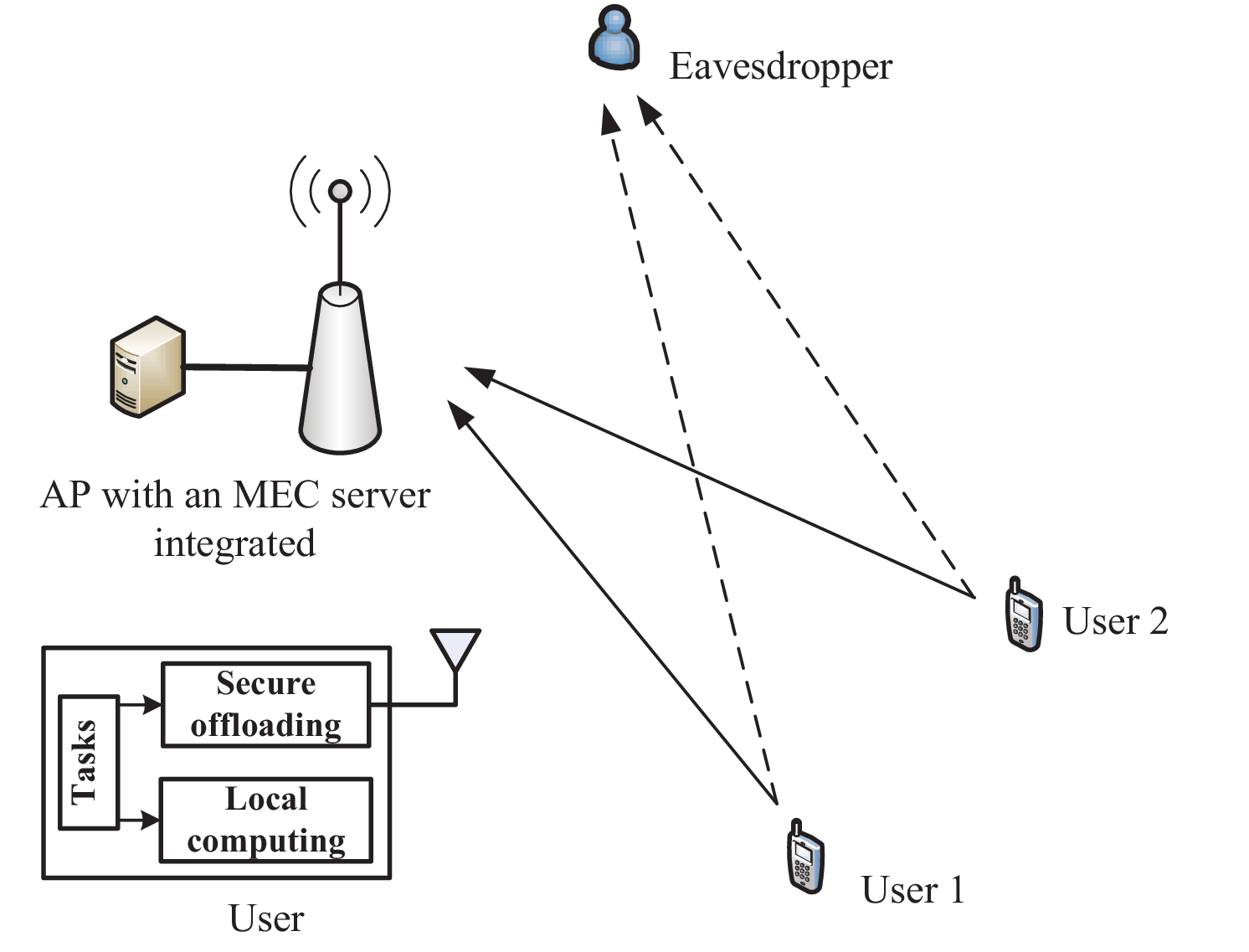}}
  \caption{The multiuser MEC system with NOMA-assisted secure computation offloading in the presence of an eavesdropper. In order to gain an insightful understanding of the uplink secure NOMA-MEC system, we focus on the fundamental two-scheduled-user case, i.e., $K=2$.}\label{fig:system}}
\end{figure}

As shown in Fig. \ref{fig:system}, an uplink NOMA communication system is considered, where $K > 1$ users can offload their computation-intensive tasks to one AP (with an MEC sever integrated) in the presence of an external eavesdropper. All the nodes are equipped with a single antenna.$\footnote{Note that the multi-antenna NOMA-MEC scenario has attracted much attention recently \cite{IEEEhowto:4, IEEEhowto:39}. In this paper, we focus on the simple single antenna case for the purpose of gaining an insightful understading of the uplink secure NOMA-MEC offloading. Moreover, to the best of our knowledge, this is the first work considering secrecy outage probability in the NOMA-based MEC networks.}$ For the ease of presentation, we utilize user $k$ to denote the $k$th user with $k \in \left\{ {1,{\kern 1pt} {\kern 1pt}  \cdots ,{\kern 1pt} {\kern 1pt} K} \right\}$. As for the wireless channels, the frequency non-selective quasi-static block fading model \cite{IEEEhowto:21} is adopted such that the channels remain unchanged during the given transmission block of our interest with a finite duration $T$. The channel coefficients from user $k$ to the AP and the eavesdropper are denoted by ${h_{AP,k}} = d_{AP,k}^{ - {\alpha  \mathord{\left/
 {\vphantom {\alpha  2}} \right.
 \kern-\nulldelimiterspace} 2}}{g_{AP,k}}$  and ${h_{e,k}} = d_{e,k}^{ - {\alpha  \mathord{\left/
 {\vphantom {\alpha  2}} \right.
 \kern-\nulldelimiterspace} 2}}{g_{e,k}}$, respectively, where ${d_{AP,k}}$  and  ${d_{e,k}}$ denote the distance from user $k$ to the AP and the eavesdropper, respectively; $\alpha $ indicates the path-loss exponent; and ${g_{AP,k}},\;{g_{e,k}} \sim \mathcal{CN}\left( {0,\;1} \right)$  are the normalized Rayleigh fading channel states. Assuming that the AP knows perfectly the instantaneous channel gain of each user, i.e., ${\left| {{h_{AP,k}}} \right|^2}$, and thus accurately knows the computation information. But it  only knows the average channel gain of the eavesdropper over different fading realizations, i.e., $ \mathbb{E}\left\{ {{{\left| {{h_{e,k}}} \right|}^2}} \right\} = d_{e,k}^{ - \alpha }$. This assumption has been widely adopted in the existing literatures  \cite{IEEEhowto:21, IEEEhowto:22, IEEEhowto:23} and the references therein. In practice, the channel statistics of the eavesdropper can be estimated per the knowledge of the fading environment, e.g., the Rayleigh fading model for the rich-scattering environment, and the distance of the eavesdropper, e.g., the $d_{e,k}^{ - \alpha }$ here \cite{IEEEhowto:19, IEEEhowto:21}.

\subsection{NOMA-Based Partial Offloading in the Presence of Eavesdropping}

In OMA-MEC, each user is typically allocated with dedicated time/frequency resource for offloading its task to the MEC server \cite{IEEEhowto:11, IEEEhowto:20}. In our considered system, by using the principle of NOMA, all the users can offload their tasks simultaneously over the same time and frequency resources. Within the block of duration $T$, each user $k$ should execute a computation task with total ${L_k} > 0$  input bits. We consider the partial offloading mode, in which the task of user $k$ can be arbitrarily partitioned into two parts with ${\ell_k}$ input bits computed locally and  ${L_k} - {\ell_k}$ input bits securely offloaded to the AP, where $0 \le {\ell_k} \le {L_k}$.$\footnote{Note that the MEC server and the AP usually have sufficient large computation ability and high transmit power, respectively. Hence, we ignore the computation time consumed at the MEC server and the downloading time for computing results sending back to the users (see, e.g., \cite{IEEEhowto:4, IEEEhowto:9, IEEEhowto:20}). However, when the MEC server's computation ability is limited, the computation time at the MEC server cannot be neglected. This case is another interesting scenario which has been studied in \cite{IEEEhowto:36} and \cite{IEEEhowto:37}.}$

To reduce the system complexity, it is further assumed that two users, namely, user $m$ and user $n$, are served at the same resource block. Thus user $n$ is admitted to the time slot $T$ which would be solely occupied by user $m$ in the OMA counterpart. The reasons why we focus on the fundamental two-scheduled-user case are from three aspects. First, the scenario of two users to perform non-orthogonal multiple access (NOMA) jointly is of practical interest. Since a NOMA system is strongly interference limited, a large number of users perform NOMA is always not realistic. Second, though SIC can be utilized to suppress the interference of multi-user NOMA, it will bring high hardware complexity to the small size low-power mobile terminals when a large number of users perform NOMA. Moreover, the signal processing latency will be further increased due to the heavier interference burden of the mobile terminals. Last, the simplified two-scheduled-user case can help us gain the fundamental and insightful understanding of the multi-user NOMA assisted secure MEC system, since the studies of multi-user NOMA schemes can be done by grouping two users together to perform NOMA jointly with the technique of user pairing (see, e.g., \cite{IEEEhowto:1}). 

The main challenges for our considered two users NOMA assisted secure MEC system are from the following two aspects.

\begin{itemize}
  \item	Since we consider the practical scenario where the transmitter does not know the eavesdropper's instantaneous channel state information, how to ensure offloading security under the assumption of knowing the statistics of the eavesdropper's channel is a challenging issue. In this case, the criteria used to measure the secrecy performance of the system is the secrecy outage probability which makes the joint design of communication and computation resource allocation much more difficult.
  \item	To gain insightful understanding of our considered system, we need to obtain closed-form expressions for the solutions of the optimization variables through solving the complicated non-convex optimization problems. This is also a challenge issue that we need to address.
\end{itemize}

The  received signals at the AP and at the eavesdropper are, respectively, given by
\begin{equation}\label{eq:F1}
{y_{AP}} = \sum\limits_{k = m,n} {\sqrt {{p_k}} {h_{AP,k}}{s_k}}  + {n_{AP}},
\end{equation}
\begin{equation}\label{eq:F2}
{y_e} = \sum\limits_{k = m,n} {\sqrt {{p_k}} {h_{e,k}}{s_k}}  + {n_e},
\end{equation}
where ${s_k} \in \mathbb{C}$ is the task-bearing signal for offloading by user $k$ with $ \mathbb{E}\left[ {{{\left| {{s_k}} \right|}^2}} \right] = 1$, and ${p_k} > 0$ is the associated transmit power, ${n_{AP}}$ and ${n_e}$ are the zero-mean AWGN at the AP with variance $\sigma _{AP}^2$ and zero-mean AWGN at the eavesdropper with variance $\sigma _e^2$, respectively.

Without loss of generality, the channel state information of two users is sorted as $|{h_{AP,m}}|<|{h_{AP,n}}|$. Due to the mechanism of uplink NOMA, similar to \cite{IEEEhowto:2, IEEEhowto:14, IEEEhowto:36}, the AP is able to perform SIC to decode the received messages and the SIC decoding order is assumed as the decreasing order of channel gains. Specifically, for the AP, it first decodes the information of user $n$ and then decodes the information of user $m$. Moreover, note that admitting user $n$ to the dedicated time slot of user $m$ should not cause any performance degradation to user $m$ ideally. The transmit powers for user $n$ and user $m$ are selected in the way that user $n$'s message is received in a lower power than user $m$'s message. Then user $n$'s message is preferred to be decoded before user $m$'s at the MEC server. Thus, the received SINRs at the AP to decode user $n$'s and user $m$'s messages are, respectively, given by
\begin{equation}\label{eq:F3}
{\Gamma_{AP,n}} = \frac{{{\gamma _{AP,n}}{p_n}}}{{1 + {\gamma _{AP,m}}{p_m}}},
\end{equation}
\begin{equation}\label{eq:F4}
{\Gamma_{AP,m}} = {\gamma _{AP,m}}{p_m},
\end{equation}
where ${\gamma _{AP,n}}{\rm{ = }}\frac{{{{\left| {{h_{AP,n}}} \right|}^2}}}{{\sigma _{AP}^2}}$ and ${\gamma _{AP,m}}{\rm{ = }}\frac{{{{\left| {{h_{AP,m}}} \right|}^2}}}{{\sigma _{AP}^2}}$.

Based on the idea of worst-case assumption, we assume that the eavesdropper can cancel the uplink user interference before decoding the information of the UL users. Thus, the received SINR at the eavesdropper of the message ${s_k}$  is given by
\begin{equation}\label{eq:F5}
{\Gamma_{e,k}} = {\gamma _{e,k}}{p_k},\;k \in \left\{ {m,\;n} \right\},
\end{equation}
where ${\gamma _{e,k}} = \frac{{{{\left| {{h_{e,k}}} \right|}^2}}}{{\sigma _e^2}}$. Note that the assumption made here overestimates the eavesdropper's ability. From the perspective of the legitimate receiver (i.e., AP), such an assumption is the so-called worst-case assumption to ensure the conservative task offloading security since the AP neither knows the eavesdropper's ability nor the instantaneous CSI. This assumption has also been employed in the previous study on the secure resource allocation of full-duplex (FD) NOMA systems \cite{IEEEhowto:25}. Under this assumption, we obtain the lower bound on the achievable system secrecy rate due to the unfavourable scenario regarding the eavesdropping capacity. Recent research shows that actually the inter-user interference can help enhance information transmission security of NOMA networks in the presence of passive eavesdropper \cite{IEEEhowto:38}. How it affects the secrecy performance of our considered NOMA-assisted MEC networks is an interesting direction to pursue in the future work.

The consumed energy of each user is from two parts,  with one from offloading its computation tasks to the MEC server and the other one from the circuit power consumption. Thus, the total energy consumption can be given as \cite{IEEEhowto:33}
\begin{equation}\label{eq:F6}
E_k^{off} = \left( {{p_k} + {p_{c,k}}} \right)T,\;\;\quad k \in \left\{ {m,\;n} \right\},
\end{equation}
where ${p_{c, k}>0}$ is the constant circuit power of user $k$.

\subsection{Local Computing at Users}

For the local computing, let ${c_k}$ denote the number of CPU cycles required for computing one task-input bit at user $k$, where $k \in \left\{ {m,\;n} \right\}$. Hence, the total number of CPU cycles required for computing ${{\ell}_k}$ input bits is ${c_k}{{\ell}_k}$. For each cycle $i \in \left\{ {1,\, \ldots ,\,{c_k}{{\ell}_k}} \right\}$, user $k$ can adjust the CPU frequency ${f_{k,i}}$ by adopting the dynamic voltage and frequency scaling (DVFS) technique \cite{IEEEhowto:3} to control the energy consumption. Therefore, the total execution time used for local computing of user $k$ is $\sum\nolimits_{i = 1}^{{c_k}{{\ell}_k}} {\frac{1}{{{f_{k,i}}}}} $. Since the local computing must be accomplished before the end of one block, we have the following computation latency constraints:
\begin{equation}\label{eq:F7}
\sum\limits_{i = 1}^{{c_k}{\ell_k}} {\frac{1}{{{f_{k,i}}}}}  \le T,\quad \forall k \in \left\{ {m,\;n} \right\}.
\end{equation}

The consumed energy of user $k$ for local computing can be expressed as a function of CPU frequency given as $E_k^{loc} = \sum\limits_{i = 1}^{{c_k}{\ell_k}} {{\varsigma _k}f_{k,i}^2} $, where ${\varsigma _k} > 0$ is the effective capacitance coefficient that depends on the chip architecture of user $k$. According to the lemma proposed in \cite{IEEEhowto:9}, since the expressions $\sum\limits_{i = 1}^{{c_k}{{\ell}_k}} {{1 \mathord{\left/
 {\vphantom {1 {{f_{k,i}}}}} \right.
 \kern-\nulldelimiterspace} {{f_{k,i}}}}} $ and $\sum\limits_{i = 1}^{{c_k}{{\ell}_k}} {{\varsigma _k}f_{k,i}^2} $ are both convex with respective to the CPU frequency $ {{f_{k,i}}} $, the best solution for minimizing the energy consumption $E_k^{loc}$ while meeting the computation latency $T$ should be that $\left\{ {{f_{k,i}}} \right\}$ are identical over different CPU cycles, which is given by
\begin{equation}\label{eq:F8}
{f_{k,1}} =  \ldots  = {f_{k,{c_k}{\ell_k}}} = {{{c_k}{\ell_k}} \mathord{\left/
 {\vphantom {{{c_k}{\ell_k}} T}} \right.
 \kern-\nulldelimiterspace} T},\quad \forall k \in \left\{ {m,\;n} \right\}.
\end{equation}

Therefore, the energy consumption $E_k^{loc}$ can be rewritten as \cite{IEEEhowto:34}
\begin{equation}\label{eq:F9}
E_k^{loc} = \frac{{{\varsigma _k}c_k^3\ell_k^3}}{{{T^2}}},\quad \forall k \in \left\{ {m,\;n} \right\}.
\end{equation}

\subsection{Secure Encoding}

We use the widely-adopted Wyner's secrecy encoding scheme \cite{IEEEhowto:15} to secure the UL information offloading. In particular, the redundant information is introduced as the rate cost to provide offloading secrecy against the eavesdropper. With this, two rate parameters for offloading data of each user $k$, namely, the codeword transmission rate, ${R_{t,k}}$ (in bits/sec/Hz), and the confidential data rate, ${R_{s,k}}$ (in bits/sec/Hz), are employed. Thus, the redundant information rate ${R_{e,k}}$ (in bits/sec/Hz) of user $k$ can be calculated as the positive rate difference ${R_{e,k}} = {R_{t,k}} - {R_{s,k}}$. The adaptive secure offloading scheme is considered in our system, such that the rate parameters ${R_{t,k}}$ and ${R_{s,k}}$ can be adaptively adjusted according to the instantaneous CSI of ${{h_{AP,k}}}$.

Since the eavesdropper's instantaneous CSI is unknown at each user, perfect security is impossible. Therefore, the secrecy outage probability is introduced to measure the secrecy performance of the task offloading \cite{IEEEhowto:21, IEEEhowto:22, IEEEhowto:28}. And the secrecy outage probability of message ${s_k}$ is expressed as \cite{IEEEhowto:28}
\begin{equation}\label{eq:F10}
{{\rm{P}}_{so,k}} = \Pr\left\{ {{R_{t,k}} - {R_{s,k}} < {C_{e,k}}} \right\},\quad \forall k \in \left\{ {m,\;n} \right\},
\end{equation}
where ${C_{e,k}} = {\log _2}\left( {1 + {\Gamma_{e,k}}} \right)$ denotes the eavesdropper's channel capacity to decode message ${s_k}$. For user $k$, if ${C_{e,k}}$ exceeds ${R_{t,k}} - {R_{s,k}}$, the offloaded data can be decoded by the eavesdropper, and a secrecy outage event, whose probability is defined in (\ref{eq:F10}), will occur.

\section{Weighted Sum-Energy Consumption Minimization}

\subsection{Problem Formulation}

Under the above setup, in this section, we pursue an energy-efficient NOMA-MEC design by focusing on the weighted sum-energy consumption minimization at the uplink users while ensuring the successful latency-constrained computation task execution and offloading security. To this end, we jointly optimize the numbers of locally computed bits ${{\ell}_k}$, the power allocation ${p_k}$, the codeword transmission rates  ${R_{t,k}}$ and the confidential data rates ${R_{s,k}}$ of the uplink users.

Mathematically, the weighted sum energy consumption minimization problem is formulated as
\begin{subequations}\label{eq:F11}
\begin{equation}\label{eq:F11a}
\left( {{\mathbf{P1}}} \right):\quad \mathop {\min }\limits_{{\bm{\ell}},\,{\bf{p}},\,{{\bf{R}}_t},\,{{\bf{R}}_s}} \sum\limits_{k = m,\,n} {{\alpha _k}\left( {{{{\varsigma _k}c_k^3\ell_k^3} \mathord{\left/
 {\vphantom {{{\varsigma _k}c_k^3\ell_k^3} {{T^2}}}} \right.
 \kern-\nulldelimiterspace} {{T^2}}} + {p_k}T} \right)}
\end{equation}
\begin{equation}\label{eq:F11b}
\quad \quad \quad s.t. {\kern 1pt} {\kern 1pt} {\kern 1pt} {\kern 1pt} {\kern 1pt} {\kern 1pt} {\kern 1pt} {\kern 1pt} {\kern 1pt} {\kern 1pt} {\kern 1pt} {\kern 1pt} {\kern 1pt} {\kern 1pt} {\kern 1pt} {\kern 1pt} {\kern 1pt} {\kern 1pt} {\kern 1pt} BT{R_{s,k}} \ge {L_k} - {\ell_k},{\kern 1pt} {\kern 1pt} {\kern 1pt} {\kern 1pt} {\kern 1pt} {\kern 1pt} {\kern 1pt} {\kern 1pt} {\kern 1pt} {\kern 1pt} \forall k \in \left\{ {m,\;n} \right\},
\end{equation}
\begin{equation}\label{eq:F11c}
 {\kern 1pt} {\kern 1pt} {\kern 1pt} {\kern 1pt} {\kern 1pt} {\kern 1pt} {\kern 1pt} {\kern 1pt} {\kern 1pt} {\kern 1pt} {\kern 1pt} {\kern 1pt} {\kern 1pt} {\kern 1pt} {\kern 1pt} {\kern 1pt} {\kern 1pt} {\kern 1pt} {\kern 1pt} {\kern 1pt} {\kern 1pt} {\kern 1pt} {\kern 1pt} {\kern 1pt} {\kern 1pt} {\kern 1pt} {\kern 1pt} {\kern 1pt} {\kern 1pt} {\kern 1pt} {\kern 1pt} \quad \quad {R_{t,k}} \le {C_{AP,k}},{\kern 1pt} {\kern 1pt} {\kern 1pt} {\kern 1pt} {\kern 1pt} {\kern 1pt} {\kern 1pt} {\kern 1pt} {\kern 1pt} \forall k \in \left\{ {m,\;n} \right\},
\end{equation}
\begin{equation}\label{eq:F11d}
 {\kern 1pt} {\kern 1pt} {\kern 1pt} {\kern 1pt} {\kern 1pt} {\kern 1pt} {\kern 1pt} {\kern 1pt} {\kern 1pt} {\kern 1pt} {\kern 1pt} {\kern 1pt} {\kern 1pt} {\kern 1pt} {\kern 1pt} {\kern 1pt} {\kern 1pt} {\kern 1pt} {\kern 1pt} {\kern 1pt} {\kern 1pt} {\kern 1pt} {\kern 1pt} {\kern 1pt} {\kern 1pt} {\kern 1pt} {\kern 1pt} {\kern 1pt} {\kern 1pt} {\kern 1pt} {\kern 1pt} {\kern 1pt} {\kern 1pt} \quad \quad  {R_{s,k}} \le {R_{t,k}},{\kern 1pt} {\kern 1pt} {\kern 1pt} {\kern 1pt} {\kern 1pt} {\kern 1pt} {\kern 1pt} {\kern 1pt} {\kern 1pt} \forall k \in \left\{ {m,\;n} \right\},
\end{equation}
\begin{equation}\label{eq:F11e}
 {\kern 1pt} {\kern 1pt} {\kern 1pt} {\kern 1pt} {\kern 1pt} {\kern 1pt} {\kern 1pt} {\kern 1pt} {\kern 1pt} {\kern 1pt} {\kern 1pt} {\kern 1pt} {\kern 1pt} {\kern 1pt} {\kern 1pt} {\kern 1pt} {\kern 1pt} {\kern 1pt} {\kern 1pt} {\kern 1pt} {\kern 1pt} {\kern 1pt} {\kern 1pt} {\kern 1pt} {\kern 1pt} {\kern 1pt} {\kern 1pt} {\kern 1pt} {\kern 1pt} {\kern 1pt} {\kern 1pt} {\kern 1pt} {\kern 1pt} {\kern 1pt} {\kern 1pt} {\kern 1pt} \quad \quad {{\rm{P}}_{so,k}} \le \varepsilon ,{\kern 1pt} {\kern 1pt} {\kern 1pt} {\kern 1pt} {\kern 1pt} {\kern 1pt} {\kern 1pt} {\kern 1pt} \forall k \in \left\{ {m,\;n} \right\},
\end{equation}
\begin{equation}\label{eq:F11f}
{\kern 1pt} {\kern 1pt} {\kern 1pt} {\kern 1pt} {\kern 1pt} {\kern 1pt} {\kern 1pt} {\kern 1pt} {\kern 1pt} {\kern 1pt} {\kern 1pt} {\kern 1pt} {\kern 1pt} {\kern 1pt} {\kern 1pt} {\kern 1pt} {\kern 1pt} {\kern 1pt} {\kern 1pt} {\kern 1pt} {\kern 1pt} {\kern 1pt} {\kern 1pt} {\kern 1pt} {\kern 1pt} {\kern 1pt}  \quad \quad {\kern 1pt} {\kern 1pt} 0 \le {\ell_k} \le \ell_k^{\max },\;{p_k} \ge 0,\;{\kern 1pt} {\kern 1pt} {\kern 1pt} {\kern 1pt} {\kern 1pt} {\kern 1pt} \forall k \in \left\{ {m,\;n} \right\},
\end{equation}
\end{subequations}
where ${\bm{\ell}} = \left[ {{{\ell}_m},\;{{\ell}_n}} \right]$ denotes the task partition vector, ${\textbf{p}} = \left[ {{p_m},\;{p_n}} \right]$ denotes the power allocation vector, ${{\bf{R}}_t} = \left[ {{R_{t,m}},\;{R_{t,n}}} \right]$ denotes the codeword transmission rate vector and ${{\bf{R}}_s} = \left[ {{R_{s,m}},\;{R_{s,\,n}}} \right]$ denotes the confidential data rate vector, ${\alpha _k} > 0$ denotes the energy weight for each user $k$, $B$ denotes the system bandwidth, ${C_{AP,k}} = {\log _2}\left( {1 + {\Gamma_{AP,k}}} \right)$ denotes the channel capacity of the AP to decode the message ${s_k}$, $0 < \varepsilon  < 1$ denotes the maximum tolerable secrecy outage probability, and ${\ell}_k^{\max }$ denotes the maximum allowable numbers of locally computed bits, which is strictly limited by both the maximum CPU frequency of user $k$ and the computing latency \cite{IEEEhowto:20}. Note that the term ${p_{c,k}}T$ is not included in (\ref{eq:F11a}) since it is a constant. Constraint (\ref{eq:F11b}) implies that the offloading rate of the NOMA transmission is characterized by the confidential data rate ${R_{s,k}}$ of each user $k$, such that the $({L_k} - {{\ell}_k})$ part of the task can be securely offloaded in $T$ time slot with bandwidth $B$. We believe that the adopted confidential data rate is more suitable than the codeword transmission rate ${R_{t,k}}$ to capture the actual task offloading rate since the codeword transmission rate is the sum rate of the offloaded task and the redundancy to provide secrecy while the actual task offloading rate is the confidential data rate. Constraint (\ref{eq:F11c}) ensures that the message ${s_k}$ can be decoded by the AP without error. The secrecy constraint (\ref{eq:F11e}) presets the maximum tolerable secrecy outage probability $\varepsilon $ for each message.

Note that due to the non-convex nature of constraints (\ref{eq:F11c}) and (\ref{eq:F11e}), problem ($\mathbf{P1}$) is undoubtedly non-convex in its current form. In the following subsection, we will find a well-structured optimal solution based on the analysis and transformation of problem ($\mathbf{P1}$). The feasibility of problem ($\mathbf{P1}$) will be studied at the end of this section. Without loss of generality, in the rest of this paper, we assume that problem ($\mathbf{P1}$) is feasible, unless stated otherwise.

\subsection{Optimal Solution to Problem ($\mathbf{P1}$)}

In this subsection, we will provide the optimal value of the decision variables ${\bf{p}}$, ${{\bf{R}}_t}$ and ${{\bf{R}}_s}$ in semi-closed forms. Firstly, Lemma \ref{lem:1} is presented as follows.

\newtheorem{Lemma}{\textit{\underline{Lemma}}}
\begin{Lemma}\label{lem:1}
The optimal solution of the decision variables ${\bm{\ell}}$, ${\bf{p}}$, ${{\bf{R}}_t}$ and ${{\bf{R}}_s}$ of problem ($\mathbf{P1}$) should satisfy

\begin{equation}\label{eq:F12}
{R_{t,k}} = \left\{ {\begin{array}{*{20}{c}}
{{{\log }_2}\left( {1 + \frac{{{\gamma _{AP,n}}{p_n}}}{{1 + {\gamma _{AP,m}}{p_m}}}} \right),\quad k = n,}\\
{{{\log }_2}\left( {1 + {\gamma _{AP,m}}{p_m}} \right),\quad k = m,}
\end{array}} \right.
\end{equation}
\begin{equation}\label{eq:F13}
BT{R_{s,k}} = {L_k} - {\ell_k},\quad \forall k \in \left\{ {m,\;n} \right\}.
\end{equation}
\end{Lemma}
\begin{IEEEproof}
We prove this lemma via contradiction. Denoting the jointly optimal values of $\left\{ {{{\ell}_k}} \right\}$, $\left\{ {{p_k}} \right\}$, $\left\{ {{R_{t,k}}} \right\}$ and $\left\{ {{R_{s,k}}} \right\}$ as $\left( \left\{ {{\ell}_k^*} \right\},\;\left\{ {p_k^*} \right\},\;\left\{ {R_{t,k}^*} \right\},\;\left\{{R_{s,k}^*} \right\} \right)$, in regard to (\ref{eq:F11c}), we assume $R_{t,n}^* < {\log _2}\left( {1 + \frac{{{\gamma _{AP,n}}p_n^*}}{{1 + {\gamma _{AP,m}}p_m^*}}} \right)$ and $R_{t,m}^* < {\log _2}( 1 + {\gamma _{AP,m}}p_m^* )$. One can find that the objective function (\ref{eq:F11a}) and the constraints (\ref{eq:F11e}) are also related to variable $ {{p_k}} $. In particular, the objective value decreases with $ {{p_k}} $, and the probability of the secrecy outage events decreases with ${p_m}$ and ${p_n}$, respectively, since the eavesdropper's channel capacity ${C_{e,k}}$ reduces with ${p_k}$, for $k \in \left\{ {m,\;n} \right\}$. Hence, there must exist another resource allocation $\left\{ {{\ell}_k^*} ,\; {{p'_k}} ,\; {R_{t,k}^*} ,\; {R_{s,k}^*}  \right\}$, where ${p'_k} = p_k^* - {\tau _k},\;\forall k \in \left\{ {m,\;n} \right\}$ and ${\tau _k}$ is a small positive value, such that $R_{t,n}^* < {\log _2}\left( {1 + \frac{{{\gamma _{AP,n}}{p'_n}}}{{1 + {\gamma _{AP,m}}{p'_m}}}} \right)$, $R_{t,m}^* < {\log _2}\left( {1 + {\gamma _{AP,m}}{p'_m}} \right)$ and the secrecy outage constraint (\ref{eq:F11e}) still hold while the objective value is further reduced. It is proved that the assumption we made above is incorrect, the equations $R_{t,k}^* = {\log _2}\left( {1 + \frac{{{\gamma _{AP,n}}p_n^*}}{{1 + {\gamma _{AP,m}}p_m^*}}} \right)$ for $k = n$ and $R_{t,k}^* = {\log _2}\left( {1 + {\gamma _{AP,m}}p_m^*} \right)$ for $k = m$ must be held in the optimal solution. Here, we complete the proof of (\ref{eq:F12}).

Similarly, with respect to (\ref{eq:F11b}), if $BTR_{s,k}^* > {L_k} - \ell_k^*,\;\forall k \in \left\{ {m,\;n} \right\}$ holds, then we can find another task partition ${{\ell}'_k} = \ell_k^* - \tau ',\;\forall k \in \left\{ {m,\;n} \right\}$, where $\tau '$ is a small positive value, such that $BTR_{s,k}^* > {L_k} - {\ell'_k},\;\forall k \in \left\{ {m,\;n} \right\}$ still holds while the objective value is further reduced. Based on this, we conclude that $BTR_{s,k}^* = {L_k} - \ell_k^*,\;\forall k \in \left\{ {m,\;n} \right\}$ must be held. Thus, we complete the proof of (\ref{eq:F13}). Lemma \ref{lem:1} is thus proved.
\end{IEEEproof}

\textit{\underline{Remark} 1:} From the perspective of optimization design, for problem $\left( {{\mathbf{P1}}} \right)$, it is easy to note that the optimal $R_{t,k}$ is the maximum $R_{t,k}$ that satisfies (\ref{eq:F11c}). This result is clearly consistent with the conclusion provided in Lemma \ref{lem:1}. Moreover, it can be seen from Lemma \ref{lem:1} that, given the secrecy outage performance of task offloading, it is an energy-efficient way to set the codeword transmission rate, $R_{t,k}$, equal to the channel capacity of user $k$ for decoding its own message.

Lemma \ref{lem:1} also provides the important insights of the relationship among the decision variables ${\bm{\ell}}$ , ${\bf{p}}$, ${{\bf{R}}_t}$ and ${{\bf{R}}_s}$. Based on the conclusion in Lemma \ref{lem:1}, we have the following theorem to reformulate the non-convex secrecy outage probability constraint (\ref{eq:F11e}).

\newtheorem{Theorem}{\textit{\underline{Theorem}}}
\begin{Theorem}\label{the:1}
Based on Lemma \ref{lem:1}, we have the following reformulation about the non-convex secrecy outage probability constraint (\ref{eq:F11e}), which is given as
\begin{subequations}\label{eq:F14}
\begin{equation}\label{eq:F14a}
{\frac{{\frac{{1 + {\gamma _{AP, m}}{p_m} + {\gamma _{AP, n}}{p_n}}}{{1 + {\gamma _{AP, m}}{p_m}}} - {2^{{R_{s, n}}}}}}{{{2^{{R_{s, n}}}}{p_n}}} \ge {a_1},}
\end{equation}
\begin{equation}\label{eq:F14b}
{\frac{{1 + {\gamma _{AP, m}}{p_m} - {2^{{R_{s, m}}}}}}{{{2^{{R_{s, m}}}}{p_m}}} \ge {a_2}},
\end{equation}
\end{subequations}
where ${a_1} = {{\ln \left( {{\varepsilon ^{ - 1}}} \right)} \mathord{\left/
 {\vphantom {{\ln \left( {{\varepsilon ^{ - 1}}} \right)} {\left( {\sigma _e^2d_{e,n}^\alpha } \right)}}} \right.
 \kern-\nulldelimiterspace} {\left( {\sigma _e^2d_{e,n}^\alpha } \right)}}$ and ${a_2} = {{\ln \left( {{\varepsilon ^{ - 1}}} \right)} \mathord{\left/
 {\vphantom {{\ln \left( {{\varepsilon ^{ - 1}}} \right)} {\left( {\sigma _e^2d_{e,m}^\alpha } \right)}}} \right.
 \kern-\nulldelimiterspace} {\left( {\sigma _e^2d_{e,m}^\alpha } \right)}}$.
\end{Theorem}

\begin{IEEEproof}
Please refer to Appendix A.
\end{IEEEproof}

With (\ref{eq:F12}), (\ref{eq:F13}) and (\ref{eq:F14}), as well as the fact that (\ref{eq:F11d}) always can be satisfied if (\ref{eq:F11e}) is satisfied, problem ($\mathbf{P1}$) is simplified and  equivalently reconstructed as
\begin{subequations}\label{eq:F18}
\begin{equation}\label{eq:F18a}
\left( {{\mathbf{P1.1}}} \right):\quad \mathop {\min }\limits_{{\bm{\ell}},\,{\bf{p}},\,\,{{\bf{R}}_s}} \sum\limits_{k = m,n} {{\alpha _k}\left( {{{{\varsigma _k}c_k^3\ell_k^3} \mathord{\left/
 {\vphantom {{{\varsigma _k}c_k^3\ell_k^3} {{T^2}}}} \right.
 \kern-\nulldelimiterspace} {{T^2}}} + {p_k}T} \right)}
\end{equation}
\begin{equation}\label{eq:F18b}
\quad \quad \quad \;s.t. {\kern 1pt} {\kern 1pt} {\kern 1pt} {\kern 1pt} {\kern 1pt} {\kern 1pt} {\kern 1pt} {\kern 1pt} {\kern 1pt} {\kern 1pt} {\kern 1pt} {\kern 1pt} {\kern 1pt} {\kern 1pt} {\kern 1pt} {\kern 1pt} {\kern 1pt} {\kern 1pt} {\kern 1pt} BT{R_{s,k}} = {L_k} - {\ell_k},{\kern 1pt} {\kern 1pt} {\kern 1pt} {\kern 1pt} {\kern 1pt} {\kern 1pt} {\kern 1pt} {\kern 1pt} \forall k \in \left\{ {m,\;n} \right\},
\end{equation}
\begin{equation}\label{eq:F18c}
 {\kern 1pt} {\kern 1pt} {\kern 1pt} {\kern 1pt} {\kern 1pt} {\kern 1pt} {\kern 1pt} {\kern 1pt} {\kern 1pt} {\kern 1pt} {\kern 1pt} {\kern 1pt} {\kern 1pt} {\kern 1pt} {\kern 1pt} {\kern 1pt} {\kern 1pt} {\kern 1pt} {\kern 1pt} {\kern 1pt} {\kern 1pt} {\kern 1pt} {\kern 1pt} {\kern 1pt} {\kern 1pt} {\kern 1pt} {\kern 1pt} {\kern 1pt} {\kern 1pt} {\kern 1pt} {\kern 1pt} {\kern 1pt} {\kern 1pt} {\kern 1pt} {\kern 1pt} {\kern 1pt} \quad \;\;\frac{{\frac{{1 + {\gamma _{AP,m}}{p_m} + {\gamma _{AP,n}}{p_n}}}{{1 + {\gamma _{AP,m}}{p_m}}} - {2^{{R_{s,n}}}}}}{{{2^{{R_{s,n}}}}{p_n}}} \ge {a_1},
\end{equation}
\begin{equation}\label{eq:F18d}
{\kern 1pt} {\kern 1pt} {\kern 1pt} {\kern 1pt} {\kern 1pt} {\kern 1pt} {\kern 1pt} {\kern 1pt} {\kern 1pt} {\kern 1pt} {\kern 1pt} {\kern 1pt} {\kern 1pt} {\kern 1pt} {\kern 1pt} {\kern 1pt} {\kern 1pt} {\kern 1pt} {\kern 1pt} {\kern 1pt} {\kern 1pt} {\kern 1pt} {\kern 1pt} {\kern 1pt} {\kern 1pt} {\kern 1pt} {\kern 1pt} {\kern 1pt} {\kern 1pt} {\kern 1pt} {\kern 1pt} {\kern 1pt} {\kern 1pt} {\kern 1pt} {\kern 1pt} {\kern 1pt} {\kern 1pt} {\kern 1pt} {\kern 1pt} {\kern 1pt} \quad {\kern 1pt} \frac{{1 + {\gamma _{AP,m}}{p_m} - {2^{{R_{s,m}}}}}}{{{2^{{R_{s,m}}}}{p_m}}} \ge {a_2},
\end{equation}
\begin{equation}\label{eq:F18e}
{\kern 1pt} {\kern 1pt} {\kern 1pt} {\kern 1pt} {\kern 1pt} {\kern 1pt} {\kern 1pt} {\kern 1pt} {\kern 1pt} {\kern 1pt} {\kern 1pt} {\kern 1pt} {\kern 1pt} {\kern 1pt} {\kern 1pt} {\kern 1pt} {\kern 1pt} {\kern 1pt} {\kern 1pt} {\kern 1pt} {\kern 1pt} {\kern 1pt} {\kern 1pt} \quad \quad {\kern 1pt} \;{\kern 1pt} \,\;0 \le {\ell_k} \le \ell_k^{\max },\;{p_k} \ge 0,\;{\kern 1pt} {\kern 1pt} {\kern 1pt} {\kern 1pt} {\kern 1pt} {\kern 1pt} \forall k \in \left\{ {m,\;n} \right\}.
\end{equation}
\end{subequations}
Note that the problem ($\mathbf{P1.1}$) is still non-convex due to the non-convex constraints (\ref{eq:F18c}) and (\ref{eq:F18d}), where the decision variables ${p_m}$, ${p_n}$, ${R_{s,m}}$ and ${R_{s,n}}$ are coupled with each other in a complex way. Thus, the standard convex optimization solver, e.g., CVX \cite{IEEEhowto:29}, is unavailable to solve this problem directly. It is readily seen that, it is an extremely challenging task to transform these non-convex constraints into the convex ones. However, for the fixed task partition case, we can first obtain the optimal secrecy offloading rates ${\bf{R}}_s^*\left( {\bm{\ell}} \right)$, transmission powers ${{\bf{p}}^{\rm{*}}}\left( {\bm{\ell}} \right)$ and codeword transmission rates ${\bf{R}}_t^*\left( {\bm{\ell}} \right)$ in closed form, and then solve problem ($\mathbf{P1.1}$) global optimally by using two-dimensional exhaustive search over ${{\ell}_m}$ and ${{\ell}_n}$. The optimal ${\bf{R}}_s^*\left( {\bm{\ell}} \right)$, ${{\bf{p}}^{\rm{*}}}\left( {\bm{\ell}} \right)$ and ${\bf{R}}_t^*\left( {\bm{\ell}} \right)$ are summarized in the following theorem.

\begin{Theorem}\label{the:2}
For a given ${\bm{\ell}} = \left[ {{\ell_m},\;{\ell_n}} \right]$, the optimal secrecy offloading rates ${\bf{R}}_s^*\left( {\bm{\ell}} \right)$, transmission powers ${{\bf{p}}^{\rm{*}}}\left( {\bm{\ell}} \right)$ and codeword transmission rates ${\bf{R}}_t^*\left( {\bm{\ell}} \right)$ to minimize the weighted sum energy consumption in our considered secrecy NOMA-MEC system are, respectively, given by
\begin{equation}\label{eq:F19}
R_{s,k}^*\left( {{\ell_k}} \right) = \frac{{{L_k} - {\ell_k}}}{BT},\quad \forall k \in \left\{ {m,\;n} \right\},
\end{equation}
\begin{subequations}\label{eq:F20}
\begin{equation}\label{eq:F20a}
{p_m^*\left( {{\ell_m}} \right) = \frac{{{2^{R_{s,m}^*\left( {{\ell_m}} \right)}} - 1}}{{{\gamma _{AP,m}} - {a_2}{2^{R_{s,m}^*\left( {{\ell_m}} \right)}}}},}
\end{equation}
\begin{equation}\label{eq:F20b}
{p_n^*\left( {{\ell_m},{\ell_n}} \right) = \frac{{\left( {1 + {\gamma _{AP,m}}p_m^*\left( {{\ell_m}} \right)} \right)\left( {{2^{R_{s,n}^*\left( {{\ell_n}} \right)}} - 1} \right)}}{{{\gamma _{AP,n}} - \left( {1 + {\gamma _{AP,m}}p_m^*\left( {{\ell_m}} \right)} \right){a_1}{2^{R_{s,n}^*\left( {{\ell_n}} \right)}}}},}
\end{equation}
\end{subequations}
\begin{subequations}\label{eq:F21}
\begin{equation}\label{eq:F21a}
{R_{t,m}^*\left( {{\ell_m}} \right) = {{\log }_2}\left( {1 + {\gamma _{AP,m}}p_m^*\left( {{\ell_m}} \right)} \right),}
\end{equation}
\begin{equation}\label{eq:F21b}
{R_{t,n}^*\left( {{\ell_m},\;{\ell_n}} \right) = {{\log }_2}\left( {1 + \frac{{{\gamma _{AP,n}}p_n^*\left( {{\ell_m},\;{\ell_n}} \right)}}{{1 + {\gamma _{AP,m}}p_m^*\left( {{\ell_m}} \right)}}} \right).}
\end{equation}
\end{subequations}
\end{Theorem}
\begin{IEEEproof}
For any given ${\ell_m}$ and ${\ell_n}$, from (\ref{eq:F18b}), the optimal values of $R_{s,m}^*\left( {{\ell_m}} \right)$ and $R_{s,n}^*\left( {{\ell_n}} \right)$ can be obtained immediately as given in (\ref{eq:F19}).

To prove the solution in (\ref{eq:F20}), we first define two functions as \[{g_1}\left( {{p_m},\;{p_n}} \right) = \frac{{\frac{{1 + {\gamma _{AP,m}}{p_m} + {\gamma _{AP,n}}{p_n}}}{{1 + {\gamma _{AP,{\kern 1pt} m}}{p_m}}} - {2^{{R_{s,n}}}}}}{{{2^{{R_{s,n}}}}{p_n}}}\] and \[{g_2}\left( {{p_m}} \right) = \frac{{1 + {\gamma _{AP, m}}{p_m} - {2^{{R_{s, m}}}}}}{{{2^{{R_{s, m}}}}{p_m}}}.\]Then, the constraints (\ref{eq:F18c}) and (\ref{eq:F18d}) can be rewritten as ${g_1}\left( {{p_m},\;{p_n}} \right) \ge {a_1}$ and ${g_2}\left( {{p_m}} \right) \ge {a_2}$, respectively. We find that $\frac{{\partial {g_1}\left( {{p_m},\;{p_n}} \right)}}{{\partial {p_m}}} =  - \frac{{{\gamma _{AP,m}}{\gamma _{AP,n}}{p_n}}}{{{2^{{R_{s,n}}}}{p_n}{{\left( {1 + {\gamma _{AP,m}}{p_m}} \right)}^2}}} < 0$ and $\frac{{\partial {g_1}\left( {{p_m},\;{p_n}} \right)}}{{\partial {p_n}}} = \frac{{{2^{{R_{s,n}}}} - 1}}{{{2^{{R_{s,n}}}}p_n^2}} \ge 0$. Together with the fact that the smaller the values of $\left\{ {{p_k}} \right\}$ are, the better the objective value of problem ($\mathbf{P1.1}$). We are further aware that 1) for any given ${p_m}$, the minimum ${p_n}$ is obtained when constraint (\ref{eq:F18c}) is active, and 2) the minimum ${p_n}$ decreases with ${p_m}$. Moreover, since ${{\partial {g_2}\left( {{p_m}} \right)} \mathord{\left/
 {\vphantom {{\partial {g_2}\left( {{p_m}} \right)} {\partial {p_m}}}} \right.
 \kern-\nulldelimiterspace} {\partial {p_m}}} = {{\left( {{2^{{R_{s,m}}}} - 1} \right)} \mathord{\left/
 {\vphantom {{\left( {{2^{{R_{s,m}}}} - 1} \right)} {\left( {{2^{{R_{s,m}}}}p_m^2} \right)}}} \right.
 \kern-\nulldelimiterspace} {\left( {{2^{{R_{s,m}}}}p_m^2} \right)}} > 0$, we find ${g_2}\left( {{p_m}} \right)$ decreases with ${p_m}$. Therefore, based on the above overview and combined with the result in (\ref{eq:F19}), we can conclude that the optimal ${p_m}$, i.e., $p_m^*\left( {{\ell_m}} \right)$ given in (\ref{eq:F20a}), is obtained once the constraint (\ref{eq:F18d}) is active. Then, by substituting $p_m^*\left( {{\ell_m}} \right)$ in (\ref{eq:F20a}) into (\ref{eq:F18c}) and combining with the result in (\ref{eq:F19}), the optimal ${p_n}$, i.e., $p_n^*\left( {{\ell_m},\;{\ell_n}} \right)$ given in (\ref{eq:F20b}), is obtained when constraint (\ref{eq:F18c}) is active.

Finally, by combining the results achieved in Lemma \ref{lem:1} and (\ref{eq:F20}), the optimal codeword transmission rates $R_{t,m}^*\left( {{\ell_m}} \right)$ and $R_{t,n}^*\left( {{\ell_m},\;{\ell_n}} \right)$ given in (\ref{eq:F21}) can be obtained straightly. This completes the proof of Theorem \ref{the:2}.
 \end{IEEEproof}

Then, by searching ${\ell_m}$ and ${\ell_n}$ over $\left[ {0,\;\ell_m^{\max }} \right]$ and $\left[ {0,\;\ell_n^{\max }} \right]$, respectively, the problem ($\mathbf{P1.1}$) can be solved optimally, and the optimal task partitions denoted by $\ell_m^{opt}$ and $\ell_n^{opt}$ can be efficiently calculated. Thus, from Theorem \ref{the:1}, we gain the final optimal solution for the remaining decision variables, which can be denoted as ${\bf{R}}_s^{opt} = \left[ {R_{s,m}^{opt},\;R_{s,n}^{opt}} \right]$, ${{\bf{p}}^{opt}} = \left[ {p_m^{opt},\;p_n^{opt}} \right]$ and ${\bf{R}}_t^{opt} = \left[ {R_{t,m}^{opt},\;R_{t,n}^{opt}} \right]$.

\textit{\underline{Remark} 2:} It can be seen from Theorem \ref{the:2} that the larger the number of local computing bit is, the lower the secrecy offloading rates and the power consumption of users are. However, though a less energy is consumed by offloading tasks when more bits are computed locally, it is not the best choice to offload computing bits as less as possible. The energy consumption balance between local computing and task offloading should be maintained to minimize the overall energy consumption of all users. Moreover, it can be seen that the power $p_n$ varies monotonously with the power $p_m$. This indicates that user $n$'s transmission power is affected by user $m$ in the form of co-channel interference.

Note that the weighted sum-energy consumption minimization problem ($\mathbf{P1}$) can also be solved suboptimally by using the widely adopted Lagrange duality method (solve the dual function with given dual variables first, then solve dual problem via updating dual variables, see, e.g., \cite{IEEEhowto:9, IEEEhowto:11, IEEEhowto:14, IEEEhowto:20}). The detailed discussion of this suboptimal solution will be tedious, and thus is omitted here.

At the end of this section, we study the feasibility of our considered problem. Based on the above analysis and conclusions, we have the following proposition to summarize the feasibility condition of the problem ($\mathbf{P1.1}$) (i.e., problem ($\mathbf{P1}$)).

\newtheorem{Proposition}{\textit{\underline{Proposition}}}
\begin{Proposition}\label{pro:1}
The problem ($\mathbf{P1.1}$) is feasible if and only if the following problem is feasible.
\begin{subequations}\label{eq:F22}
\begin{equation}\label{eq:F22a}
\mathop {\min }\limits_{{\ell_m},{p_m},{R_{s,m}}} \;{{{\varsigma _m}c_m^3\ell_m^3} \mathord{\left/
 {\vphantom {{{\varsigma _m}c_m^3\ell_m^3} {{T^2}}}} \right.
 \kern-\nulldelimiterspace} {{T^2}}} + {p_m}T
\end{equation}
\begin{equation}\label{eq:F22b}
s.t.{\kern 1pt} {\kern 1pt} {\kern 1pt} {\kern 1pt} {\kern 1pt} {\kern 1pt} {\kern 1pt} {\kern 1pt} {\kern 1pt} {\kern 1pt} {\kern 1pt} {\kern 1pt} {\kern 1pt} {\kern 1pt} {\kern 1pt} {\kern 1pt} {\kern 1pt} {\kern 1pt} {\kern 1pt} {\kern 1pt} {\kern 1pt} {\kern 1pt} BT{R_{s,\,m}} = {L_m} - {\ell_m},{\kern 1pt}
\end{equation}
\begin{equation}\label{eq:F22c}
{\kern 1pt} {\kern 1pt} {\kern 1pt} {\kern 1pt} {\kern 1pt} {\kern 1pt} {\kern 1pt} {\kern 1pt} {\kern 1pt} {\kern 1pt} {\kern 1pt} {\kern 1pt} {\kern 1pt} {\kern 1pt} {\kern 1pt} {\kern 1pt} {\kern 1pt} {\kern 1pt} {\kern 1pt} {\kern 1pt} {\kern 1pt} {\kern 1pt} {\kern 1pt} {\kern 1pt} {\kern 1pt} {\kern 1pt} {\kern 1pt} {\kern 1pt} \;\,\frac{{1 + {\gamma _{AP,m}}{p_m} - {2^{{R_{s,m}}}}}}{{{2^{{R_{s,m}}}}{p_m}}} \ge {a_2},
\end{equation}
\begin{equation}\label{eq:F22d}
{\kern 1pt} {\kern 1pt} {\kern 1pt} {\kern 1pt} {\kern 1pt} {\kern 1pt} {\kern 1pt} {\kern 1pt} {\kern 1pt} {\kern 1pt} {\kern 1pt} {\kern 1pt} {\kern 1pt} {\kern 1pt} {\kern 1pt} {\kern 1pt} {\kern 1pt} {\kern 1pt} {\kern 1pt} {\kern 1pt} {\kern 1pt} {\kern 1pt} {\kern 1pt} {\kern 1pt} {\kern 1pt} {\kern 1pt} {\kern 1pt} {\kern 1pt} {\kern 1pt} \;0 \le {\ell_m} \le \ell_m^{\max },\quad {p_m} \ge 0.
\end{equation}
\end{subequations}
\end{Proposition}

\begin{IEEEproof}
First, it is easy to verify that if problem (\ref{eq:F22}) is infeasible, then problem ($\mathbf{P1.1}$) cannot be feasible since problem ($\mathbf{P1.1}$) contains additional constraints of user $n$. Secondly, if problem (\ref{eq:F22}) is feasible, and let $\left( {{\ell_m},\;{p_m},\;{R_{s,m}}} \right)$ be a feasible solution. Then, we can also find a new solution $\left( {{\ell_m},\;{p_m},\;{R_{s,m}},\;{\ell_n},\;{p_n},\;{R_{s,n}}} \right)$ which is also feasible for problem (\ref{eq:F22}) and satisfies the constraints of problem ($\mathbf{P1.1}$). The newly added solution $\left( {{\ell_n},\;{p_n},\;{R_{s,n}}} \right)$ plays the role in satisfying the constraints related to user $n$ of problem ($\mathbf{P1.1}$). Hence, we conclude that problem ($\mathbf{P1.1}$) is feasible. Proposition \ref{pro:1} is thus proved.
\end{IEEEproof}

Proposition \ref{pro:1} indicates that the feasibility of problem ($\mathbf{P1.1}$) can only depend on the constraints related to user $m$ and can be checked by solving problem (\ref{eq:F22}) via one-dimensional exhaustive search over the region $\left[ {0,\;{\ell_m}} \right]$. The closed form expressions for the decision variables ${p_m}$ and ${R_{s,m}}$ are provided in Theorem \ref{the:2}.

For summarizing, we present the details of our proposed optimal solution to problem ($\mathbf{P1}$) in Algorithm 1.

\begin{algorithm}[!t]    
\label{alg:Algorithm 1}     
\caption{Optimal solution to problem ($\mathbf{P1}$)}   
\begin{algorithmic}[1]
\STATE \textbf{Setting}\\ $B$, $T$, $\alpha$, $\alpha_m, \alpha_n$, $\varsigma_m, \varsigma_n$, $c_m, c_n$, $L_m, L_n$, $\varepsilon$;\\
channel condition: $\gamma_{AP,m}, \gamma_{AP,n}$ and $\gamma_{e,k}$;
\STATE \textbf{Initialization} \\ ${\ell_m}$ and ${\ell_n}$;
\STATE\textbf{Repeat}
  \STATE ~~search ${\ell_m}$ and ${\ell_n}$ via bisection method;
  \STATE ~~calculate ${\bf{p}}$, ${{\bf{R}}_t}$ and ${{\bf{R}}_s}$ through Theorem \ref{the:2};
\STATE \textbf{Until} ${\ell_m}$ and ${\ell_n}$ converge within a prescribed accuracy.
\STATE \textbf{output} \\$\ell_{m}^*, \ell_{n}^*$, $p_{m}^*, p_{n}^*$, $R_{s,m}^*, R_{s,n}^*$, $R_{t,m}^*$ and $R_{t,n}^*$.
\end{algorithmic}
\end{algorithm}

\emph{\underline{Complexity analysis}}: The complexity of Algorithm 1 mainly comes from the bisection method used for obtaining the local computing bits and the computation of the transmission powers, the secrecy offloading rates and the codeword transmission rates. Let $\xi$ denote the tolerance error for the bisection method. Note that the computation of the transmission powers, the secrecy offloading rates and the codeword transmission rates is carried out in each search step. Thus, according to the works in \cite{IEEEhowto:18} and \cite{IEEEhowto:19}, the total complexity of Algorithm 1 is $\mathcal{O}\left[6\log _2^2\left( {{\xi  \mathord{\left/
 {\vphantom {\xi  T}} \right.
 \kern-\nulldelimiterspace} T}} \right)\right]$ and $\mathcal{O}(\cdot)$ is the big-O notation.

\section{Secrecy Outage Probability Minimization}

In this section, we study the priority-based resource allocation problem in order to minimize the secrecy outage probability of both uplink users subject to the successful latency-constrained computation task execution constraints and energy budget constraints. In particular, as mentioned in Section II-A, the admitting of user $n$ to time slot $T$ should not cause any performance degradation to user $m$. Motivated by this requirement, we pursue the priority-based design such that the secrecy outage probability performance of user $m$ can receive preferred attention, while user $n$'s secrecy outage performance is the second.
\subsection{Problem Formulation}

Based on the analysis and model provided in the above sections, we formulate the secrecy outage probability minimization problem as
\begin{subequations}\label{eq:F23}
\begin{equation}\label{eq:F23a}
\left( {{\mathbf{P2}}} \right):\quad \;\;\mathop {\min }\limits_{{\bm{\ell}},\,{\bf{p}},\,{{\bf{R}}_t},\,{{\bf{R}}_s}} \left\{ {{\kern 1pt} {{\rm{P}}_{so,m}},{\kern 1pt} \;{{\rm{P}}_{so,n}}} \right\}
\end{equation}
\begin{equation}\label{eq:F23b}
~~\quad \quad \; \; s.t. {\kern 1pt} {\kern 1pt} {\kern 1pt} {\kern 1pt} {\kern 1pt} {\kern 1pt} {\kern 1pt} {\kern 1pt} {\kern 1pt} {\kern 1pt} {\kern 1pt} {\kern 1pt} {\kern 1pt} {\kern 1pt} {\kern 1pt} {\kern 1pt} {\kern 1pt} {\kern 1pt} {\kern 1pt} BT{R_{s,k}} \ge {L_k} - {\ell_k},{\kern 1pt} {\kern 1pt} {\kern 1pt} {\kern 1pt} {\kern 1pt} {\kern 1pt} {\kern 1pt} {\kern 1pt} {\kern 1pt} {\kern 1pt} \forall k \in \left\{ {m,\;n} \right\},
\end{equation}
\begin{equation}\label{eq:F23c}
{\kern 1pt} {\kern 1pt} {\kern 1pt} {\kern 1pt} {\kern 1pt} {\kern 1pt} {\kern 1pt} {\kern 1pt} {\kern 1pt} {\kern 1pt} {\kern 1pt} {\kern 1pt} {\kern 1pt} {\kern 1pt} {\kern 1pt} {\kern 1pt} {\kern 1pt} {\kern 1pt} {\kern 1pt} {\kern 1pt} {\kern 1pt} {\kern 1pt} {\kern 1pt} {\kern 1pt} {\kern 1pt} {\kern 1pt} {\kern 1pt} {\kern 1pt} {\kern 1pt} {\kern 1pt} {\kern 1pt} {\kern 1pt} {\kern 1pt} {\kern 1pt} \quad \quad \;{R_{t,k}} \le {C_{AP,k}},{\kern 1pt} {\kern 1pt} {\kern 1pt} {\kern 1pt} {\kern 1pt} {\kern 1pt} {\kern 1pt} {\kern 1pt} {\kern 1pt} \forall k \in \left\{ {m,\;n} \right\},
\end{equation}
\begin{equation}\label{eq:F23d}
{\kern 1pt} {\kern 1pt} {\kern 1pt} {\kern 1pt} {\kern 1pt} {\kern 1pt} {\kern 1pt} {\kern 1pt} {\kern 1pt} {\kern 1pt} {\kern 1pt} {\kern 1pt} {\kern 1pt} {\kern 1pt} {\kern 1pt} {\kern 1pt} {\kern 1pt} {\kern 1pt} {\kern 1pt} {\kern 1pt} {\kern 1pt} {\kern 1pt} {\kern 1pt} {\kern 1pt} {\kern 1pt} {\kern 1pt} {\kern 1pt} {\kern 1pt} {\kern 1pt} {\kern 1pt} {\kern 1pt} {\kern 1pt} {\kern 1pt} {\kern 1pt} \quad \quad \; \,{R_{s,k}} \le {R_{t,k}},{\kern 1pt} {\kern 1pt} {\kern 1pt} {\kern 1pt} {\kern 1pt} {\kern 1pt} {\kern 1pt} {\kern 1pt} {\kern 1pt} \forall k \in \left\{ {m,\;n} \right\},
\end{equation}
\begin{equation}\label{eq:F23e}
{\kern 1pt} {\kern 1pt} {\kern 1pt} {\kern 1pt} {\kern 1pt} {\kern 1pt} {\kern 1pt} {\kern 1pt} {\kern 1pt} {\kern 1pt} {\kern 1pt} {\kern 1pt} {\kern 1pt} {\kern 1pt} {\kern 1pt} {\kern 1pt} {\kern 1pt} {\kern 1pt} {\kern 1pt} {\kern 1pt} {\kern 1pt} {\kern 1pt} {\kern 1pt} {\kern 1pt} {\kern 1pt} {\kern 1pt} {\kern 1pt} {\kern 1pt} {\kern 1pt} {\kern 1pt} {\kern 1pt} {\kern 1pt} {\kern 1pt} {\kern 1pt} \quad \;\;{{{\varsigma _k}c_k^3\ell_k^3} \mathord{\left/
 {\vphantom {{{\varsigma _k}c_k^3l_k^3} {{T^2}}}} \right.
 \kern-\nulldelimiterspace} {{T^2}}} + {p_k}T \le {E_k},{\kern 1pt} {\kern 1pt} {\kern 1pt} {\kern 1pt} {\kern 1pt} {\kern 1pt} {\kern 1pt} \forall k \in \left\{ {m,\;n} \right\},
\end{equation}
\begin{equation}\label{eq:F23f}
{\kern 1pt} {\kern 1pt} {\kern 1pt} {\kern 1pt} {\kern 1pt} {\kern 1pt} {\kern 1pt} {\kern 1pt} {\kern 1pt} {\kern 1pt} {\kern 1pt} {\kern 1pt} {\kern 1pt} {\kern 1pt} {\kern 1pt} {\kern 1pt} {\kern 1pt} {\kern 1pt} {\kern 1pt} {\kern 1pt} {\kern 1pt} {\kern 1pt} {\kern 1pt} {\kern 1pt} {\kern 1pt} {\kern 1pt} {\kern 1pt} \quad \;\;\;\;\;{\kern 1pt} 0 \le {\ell_k} \le \ell_k^{\max },\;{p_k} \ge 0,\;{\kern 1pt} {\kern 1pt} {\kern 1pt} \forall k \in \left\{ {m,\;n} \right\},
\end{equation}
\end{subequations}
where (\ref{eq:F23e}) represents the energy constraints of two users, ${E_k} > 0,\;\forall k \in \left\{ {m,\;n} \right\}$ denotes the maximum available energy budget of user $k$, and $\ell_k^{\max } < {L_k}$ must holds in the partial offloading mode.

Note that due to the non-convex nature of objective functions and constraint (\ref{eq:F23c}), problem ($\mathbf{P2}$) is undoubtedly non-convex in its current form. In the following subsection, we will find the well-structured optimal solution in closed form based on the analysis and transformation of problem ($\mathbf{P2}$). Though the min-max fairness is a good idea in solving the multi-objective optimization problems, it does not work for problem $\left( {{\mathbf{P2}}} \right)$ since the newly introduced variable for the objective functions will lead to very complex coupling among multiple variables. Hence, we do not consider the criterion of the min-max fairness for problem $\left( {{\mathbf{P2}}} \right)$.

\subsection{Optimal Solution of Problem ($\mathbf{P2}$)}

One can find that the optimal ${R_{t,k}}$ is the maximum one and the optimal ${R_{s,k}}$ is the minimum one that satisfies (\ref{eq:F23c}) such that the occurrence probability of event ${R_{t,k}} - {R_{s,k}} < {C_{e,k}}$ will be as small as possible. Therefore, the expression of ${R_{t,k}}$ and ${R_{s,\,k}}$ in problem ($\mathbf{P2}$) can be given as the same as that in Lemma \ref{lem:1}. From (\ref{eq:F23e}) and (\ref{eq:F23f}), we note that ${E_k} > \frac{{{\varsigma _k}c_k^3{{\left( {l_k^{\max }} \right)}^3}}}{{{T^2}}}$ must be hold such that there is enough energy allocated for task offloading.

As given in Theorem \ref{the:1}, combing with the conclusion about ${R_{t,k}}$, the secrecy outage probabilities of ${{\rm{P}}_{so,m}}$ and ${{\rm{P}}_{so,n}}$ can be respectively written as
\begin{equation}\label{eq:F24}
{{\rm{P}}_{so,m}} = {e^{ - {{\left[ {\left( {1 + {\gamma _{AP,m}}{p_m} - {2^{{R_{s,m}}}}} \right)\sigma _e^2d_{e,m}^\alpha } \right]} \mathord{\left/
 {\vphantom {{\left[ {\left( {1 + {\gamma _{AP,m}}{p_m} - {2^{{R_{s,m}}}}} \right)\sigma _e^2d_{e,m}^\alpha } \right]} {\left( {{2^{{R_{s,m}}}}{p_m}} \right)}}} \right.
 \kern-\nulldelimiterspace} {\left( {{2^{{R_{s,m}}}}{p_m}} \right)}}}},
\end{equation}
\begin{equation}\label{eq:F25}
{{\rm{P}}_{so,n}} = {e^{ - \frac{{ {\left[ {1 + {\gamma _{AP,m}}{p_m} + {\gamma _{AP,n}}{p_n} - \left( {1 + {\gamma _{AP,m}}{p_m}} \right){2^{{R_{s,n}}}}} \right]\sigma _e^2d_{e,n}^\alpha } }}{{ {\left( {1 + {\gamma _{AP,m}}{p_m}} \right){2^{{R_{s,n}}}}{p_n}} }}}}.
\end{equation}
The problem ($\mathbf{P2}$) is then simplified as
\begin{subequations}\label{eq:F26}
\begin{equation}\label{eq:F26a}
\left( {{\mathbf{P2.1}}} \right):\;\;\mathop {\max }\limits_{{\bm{\ell}},\,{\bf{p}},\,{{\bf{R}}_s}} \left\{ {{\kern 1pt} {x_{m}}\left( {{p_m},{\kern 1pt}{R_{s,m}}} \right), \;{x_{n}}\left( {{p_m},{\kern 1pt}{p_n},{\kern 1pt}{R_{s,n}}}\right)}\right\}
\end{equation}
\begin{equation}\label{eq:F26b}
~\quad   \;s.t. {\kern 1pt} {\kern 1pt} {\kern 1pt} {\kern 1pt} {\kern 1pt} {\kern 1pt} {\kern 1pt} {\kern 1pt} {\kern 1pt} {\kern 1pt} {\kern 1pt} {\kern 1pt} {\kern 1pt} {\kern 1pt} {\kern 1pt} {\kern 1pt} {\kern 1pt} {\kern 1pt} BT{R_{s,k}} = {L_k} - {\ell_k},{\kern 1pt} {\kern 1pt} {\kern 1pt} {\kern 1pt} {\kern 1pt} {\kern 1pt} {\kern 1pt} {\kern 1pt} {\kern 1pt} {\kern 1pt} \forall k \in \left\{ {m,\;n} \right\},
\end{equation}
\begin{equation}\label{eq:F26c}
{\kern 1pt} {\kern 1pt} {\kern 1pt} {\kern 1pt} {\kern 1pt} {\kern 1pt} {\kern 1pt} {\kern 1pt} {\kern 1pt} {\kern 1pt} {\kern 1pt} {\kern 1pt} {\kern 1pt} {\kern 1pt} {\kern 1pt} {\kern 1pt} {\kern 1pt} {\kern 1pt} {\kern 1pt} {\kern 1pt} {\kern 1pt} {\kern 1pt} {\kern 1pt} {\kern 1pt} {\kern 1pt} {\kern 1pt} {\kern 1pt} {\kern 1pt} \;\;\;\;\;{{{\varsigma _k}c_k^3\ell_k^3} \mathord{\left/
 {\vphantom {{{\varsigma _k}c_k^3\ell_k^3} {{T^2}}}} \right.
 \kern-\nulldelimiterspace} {{T^2}}} + {p_k}T \le {E_k},{\kern 1pt} {\kern 1pt} {\kern 1pt} {\kern 1pt} {\kern 1pt} {\kern 1pt} {\kern 1pt} {\kern 1pt} {\kern 1pt} {\kern 1pt} \forall k \in \left\{ {m,\;n} \right\},
\end{equation}
\begin{equation}\label{eq:F26d}
{\kern 1pt} {\kern 1pt} {\kern 1pt} {\kern 1pt} {\kern 1pt} {\kern 1pt} {\kern 1pt} {\kern 1pt} {\kern 1pt} {\kern 1pt} {\kern 1pt} {\kern 1pt} {\kern 1pt} {\kern 1pt} {\kern 1pt} {\kern 1pt} {\kern 1pt} {\kern 1pt} {\kern 1pt} {\kern 1pt} {\kern 1pt} {\kern 1pt} {\kern 1pt} {\kern 1pt} {\kern 1pt} {\kern 1pt} {\kern 1pt}  \;\;\;\;\;{\kern 1pt} \;0 \le {\ell_k} \le \ell_k^{\max },\;{p_k} \ge 0,\;{\kern 1pt} {\kern 1pt} {\kern 1pt} {\kern 1pt} {\kern 1pt} {\kern 1pt} \forall k \in \left\{ {m,\;n} \right\},
\end{equation}
\end{subequations}
where \[{x_{m}}\left( {{p_m},{\kern 1pt}{R_{s,m}}} \right) = \frac{{\left( {1 + {\gamma _{AP, m}}{p_m} - {2^{{R_{s,m}}}}} \right)\sigma _e^2d_{e,m}^\alpha }}{{{2^{{R_{s,m}}}}{p_m}}},\] \[{x_{n}}\left( {{p_m},\;{p_n},\;{R_{s,\,n}}} \right) = \frac{{\left( {1 - {2^{{R_{s,n}}}} + \frac{{{\gamma _{AP, n}}{p_n}}}{{1 + {\gamma _{AP,m}}{p_m}}}} \right)\sigma _e^2d_{e,n}^\alpha }}{{{2^{{R_{s, n}}}}{p_n}}},\] ${{\rm{P}}_{so,m}} = {e^{ - {x_{m}}\left( {{p_m},\;{R_{s,m}}} \right)}}$ and ${{\rm{P}}_{so,n}} = {e^{ - {x_{n}}\left( {{p_m},\;{p_n},\;{R_{s,n}}} \right)}}$. By analyzing the problem ($\mathbf{P2.1}$), under the mind of priority-based resource allocation, we obtain the jointly optimal task partition ${\bm{\ell}}$, power allocation ${\bf{p}}$, codeword transmission rate ${{\bf{R}}_t}$ and secrecy offloading rate ${{\bf{R}}_s}$ in the following theorem.

\begin{Theorem}\label{the:3}
Based on the setup of intrinsic priority among two users, the jointly optimal ${\bm{\ell}}$, ${\bf{p}}$, ${{\bf{R}}_s}$ and ${{\bf{R}}_t}$ of the secrecy outage probability minimization problem ($\mathbf{P2.1}$) are, respectively, given by
\begin{subequations}\label{eq:F27}
\begin{equation}\label{eq:F27a}
\ell_k^{opt} = \ell_k^{\max },\;\forall k \in \left\{ {m,\;n} \right\},
\end{equation}
\begin{equation}\label{eq:F27b}
p_k^{opt} = \frac{{{E_k} - {{\left( {{\varsigma _k}c_k^3{{\left( {\ell_k^{\max }} \right)}^3}} \right)} \mathord{\left/
 {\vphantom {{\left( {{\varsigma _k}c_k^3{{\left( {\ell_k^{\max }} \right)}^3}} \right)} {{T^2}}}} \right.
 \kern-\nulldelimiterspace} {{T^2}}}}}{T},\;\forall k \in \left\{ {m,\;n} \right\},
\end{equation}
\begin{equation}\label{eq:F27c}
R_{s,k}^{opt} = \frac{{{L_k} - \ell_k^{\max }}}{BT},\;\forall k \in \left\{ {m,\;n} \right\},
\end{equation}
\begin{equation}\label{eq:F27d}
R_{t,m}^{opt} = {\log _2}\left( {1 + {\gamma _{AP,m}}p_m^{opt}} \right),
\end{equation}
\begin{equation}\label{eq:F27e}
R_{t,n}^{opt} = {\log _2}\left( {1 + \frac{{{\gamma _{AP,n}}p_n^{opt}}}{{1 + {\gamma _{AP,m}}p_m^{opt}}}} \right).
\end{equation}
\end{subequations}
\end{Theorem}

\begin{IEEEproof}
The main processes to prove Theorem \ref{the:3} are given as follows. Since user $m$ enjoys higher priority, we first focus on the quality of service (QoS) requirement of user $m$. Then, we have the secrecy outage probability minimization problem extracted from problem ($\mathbf{P2.1}$) as
\begin{subequations}\label{eq:F28}
\begin{equation}\label{eq:F28a}
\left( {{\mathbf{P2.1-m}}} \right):\;\;\mathop {\max }\limits_{{\bm{\ell}},\,{\bf{p}},\,{{\bf{R}}_s}} {x_{m}}\left( {{p_m},\;{R_{s,m}}} \right)
\end{equation}
\begin{equation}\label{eq:F28b}
\quad \quad \quad \quad \;\;\;s.t.{\kern 1pt} {\kern 1pt} {\kern 1pt} {\kern 1pt} {\kern 1pt} {\kern 1pt} {\kern 1pt} {\kern 1pt} {\kern 1pt} {\kern 1pt} {\kern 1pt} {\kern 1pt} {\kern 1pt} {\kern 1pt} {\kern 1pt} {\kern 1pt} {\kern 1pt} {\kern 1pt} {\kern 1pt} {\kern 1pt} {\kern 1pt} {\kern 1pt} BT{R_{s,m}} = {L_m} - {\ell_m},
\end{equation}
\begin{equation}\label{eq:F28c}
\quad \quad {\kern 1pt} {\kern 1pt} {\kern 1pt} {\kern 1pt} {\kern 1pt} {\kern 1pt} {\kern 1pt} {\kern 1pt} {\kern 1pt} {\kern 1pt} {\kern 1pt} {\kern 1pt} {\kern 1pt} {\kern 1pt} {\kern 1pt} {\kern 1pt} {\kern 1pt} {\kern 1pt} {\kern 1pt} {\kern 1pt} {\kern 1pt} {\kern 1pt} {\kern 1pt} {\kern 1pt} {\kern 1pt} {\kern 1pt} {\kern 1pt} {\kern 1pt} {\kern 1pt} {\kern 1pt} {\kern 1pt} {\kern 1pt} {\kern 1pt} {\kern 1pt} \quad \;\;\;\;\quad {{{\varsigma _m}c_m^3\ell_m^3} \mathord{\left/
 {\vphantom {{{\varsigma _m}c_m^3\ell_m^3} {{T^2}}}} \right.
 \kern-\nulldelimiterspace} {{T^2}}} + {p_m}T \le {E_m},
\end{equation}
\begin{equation}\label{eq:F28d}
\quad \quad {\kern 1pt} {\kern 1pt} {\kern 1pt} {\kern 1pt} {\kern 1pt} {\kern 1pt} {\kern 1pt} {\kern 1pt} {\kern 1pt} {\kern 1pt} {\kern 1pt} {\kern 1pt} {\kern 1pt} {\kern 1pt} {\kern 1pt} {\kern 1pt} {\kern 1pt} {\kern 1pt} {\kern 1pt} {\kern 1pt} {\kern 1pt} {\kern 1pt} {\kern 1pt} {\kern 1pt} {\kern 1pt} {\kern 1pt} {\kern 1pt} {\kern 1pt} {\kern 1pt} {\kern 1pt} {\kern 1pt} {\kern 1pt} {\kern 1pt} \quad \;\;\;\;\;{\kern 1pt} \;\quad \;0 \le {\ell_m} \le \ell_m^{\max },\;{p_m} \ge 0.
\end{equation}
\end{subequations}
From Theorem \ref{the:2}, we have that ${x_{m}}\left( {{p_m},\;{R_{s,m}}} \right)$ monotonously increases with ${p_m}$. Moreover, note that ${x_{m}}\left( {{p_m},\;{R_{s,m}}} \right)$ monotonously increases as ${R_{s,m}}$ decreases. Thus, from (\ref{eq:F28b}), we can also find that ${x_{m}}\left( {{p_m},\;{R_{s,m}}} \right)$ monotonously increases with ${\ell_m}$, which indicates that the maximum objective value can be obtained when the constraint (\ref{eq:F28c}) is active. According to the above analysis, we can rewrite ${x_{m}}\left( {{p_m},\;{R_{s,m}}} \right)$ as
\[\begin{array}{l}
{x_{m}}\left( {{p_m},\;{R_{s,m}}} \right) = \frac{{\left( {{2^{{{\left( {{\ell_m} - {L_m}} \right)} \mathord{\left/
 {\vphantom {{\left( {{\ell_m} - {L_m}} \right)} T}} \right.
 \kern-\nulldelimiterspace} T}}} - 1} \right)T}}{{{E_m} - {{\left( {{\varsigma _m}c_m^3\ell_m^3} \right)} \mathord{\left/
 {\vphantom {{\left( {{\varsigma _m}c_m^3\ell_m^3} \right)} {{T^2}}}} \right.
 \kern-\nulldelimiterspace} {{T^2}}}}} + {\gamma _{AP,m}}{2^{{{\left( {{\ell_m} - {L_m}} \right)} \mathord{\left/
 {\vphantom {{\left( {{\ell_m} - {L_m}} \right)} T}} \right.
 \kern-\nulldelimiterspace} T}}}\\
~~~\quad \quad \quad \quad \quad \; = {x_{m}}\left( {{\ell_m}} \right).
\end{array}\]
From ${x_{m}}( {{\ell_m}})$, one can find that it is also a monotonously increasing function with respect to ${\ell_m}$. Hence, we have the optimal ${\ell_m}$ as $\ell_m^{opt} = \ell_m^{\max }$ to maximize the objective value of problem (P2.1-m). Next, by plugging $\ell_m^{opt}$ into the active constraints $BT{R_{s,m}} = {L_m} - {\ell_m}$, ${{{\varsigma _m}c_m^3\ell_m^3} \mathord{\left/
 {\vphantom {{{\varsigma _m}c_m^3\ell_m^3} {{T^2}}}} \right.
 \kern-\nulldelimiterspace} {{T^2}}} + {p_m}T = {E_m}$ and ${R_{t,m}} = {\log _2}\left( {1 + {\gamma _{AP,m}}{p_m}} \right)$, we obtain the jointly optimal $p_m^{opt}$, $R_{s,m}^{opt}$ and $R_{t,m}^{opt}$ as given in (\ref{eq:F27b}), (\ref{eq:F27c}) and (\ref{eq:F27d}), respectively.

Then, we focus on the optimization design of user $n$ whose secrecy outage probability minimization problem is given by
\begin{subequations}\label{eq:F29}
\begin{equation}\label{eq:F29a}
\left( {{\mathbf{P2.1-n}}} \right):\;\;\mathop {\max }\limits_{{\bm{\ell}},\,{\bf{p}},\,{{\bf{R}}_s}} {x_{n}}\left( {p_m^{opt},\;{p_n},\;{R_{s,n}}} \right)
\end{equation}
\begin{equation}\label{eq:F29b}
\quad\quad \quad \quad \quad \;\;\;s.t.{\kern 1pt} {\kern 1pt} {\kern 1pt} {\kern 1pt} {\kern 1pt} {\kern 1pt} {\kern 1pt} {\kern 1pt} {\kern 1pt} {\kern 1pt} {\kern 1pt} {\kern 1pt} {\kern 1pt} {\kern 1pt} {\kern 1pt} {\kern 1pt} {\kern 1pt} {\kern 1pt} {\kern 1pt} {\kern 1pt} {\kern 1pt} {\kern 1pt} BT{R_{s,n}} = {L_n} - {\ell_n},
\end{equation}
\begin{equation}\label{eq:F29c}
\quad \quad {\kern 1pt} {\kern 1pt} {\kern 1pt} {\kern 1pt} {\kern 1pt} {\kern 1pt} {\kern 1pt} {\kern 1pt} {\kern 1pt} {\kern 1pt} {\kern 1pt} {\kern 1pt} {\kern 1pt} {\kern 1pt} {\kern 1pt} {\kern 1pt} {\kern 1pt} {\kern 1pt} {\kern 1pt} {\kern 1pt} {\kern 1pt} {\kern 1pt} {\kern 1pt} {\kern 1pt} {\kern 1pt} {\kern 1pt} {\kern 1pt} {\kern 1pt} {\kern 1pt} {\kern 1pt} {\kern 1pt} {\kern 1pt} {\kern 1pt} {\kern 1pt} \quad \;\;\;\;\quad {{{\varsigma _n}c_n^3\ell_n^3} \mathord{\left/
 {\vphantom {{{\varsigma _n}c_n^3\ell_n^3} {{T^2}}}} \right.
 \kern-\nulldelimiterspace} {{T^2}}} + {p_n}T \le {E_n},
\end{equation}
\begin{equation}\label{eq:F29d}
\quad \quad {\kern 1pt} {\kern 1pt} {\kern 1pt} {\kern 1pt} {\kern 1pt} {\kern 1pt} {\kern 1pt} {\kern 1pt} {\kern 1pt} {\kern 1pt} {\kern 1pt} {\kern 1pt} {\kern 1pt} {\kern 1pt} {\kern 1pt} {\kern 1pt} {\kern 1pt} {\kern 1pt} {\kern 1pt} {\kern 1pt} {\kern 1pt} {\kern 1pt} {\kern 1pt} {\kern 1pt} {\kern 1pt} {\kern 1pt} {\kern 1pt} {\kern 1pt} {\kern 1pt} \quad \;\;\;\;\;{\kern 1pt} \;\quad \;0 \le {\ell_n} \le \ell_n^{\max },\;{p_n} \ge 0.
\end{equation}
\end{subequations}
It is easy to see that, with given $p_m^{opt}$, problem ($\mathbf{P2.1-n}$) can be dealt with a similar method to problem ($\mathbf{P2.1-m}$). Through analyzing the monotonicity of the objective function in (\ref{eq:F29a}) about the decision variables and the tightness of constraint (\ref{eq:F29c}), the jointly optimal solution $\ell_n^{opt}$, $p_n^{opt}$, $R_{s,n}^{opt}$ and $R_{t,n}^{opt}$ can be obtained as given in (\ref{eq:F27a}), (\ref{eq:F27b}), (\ref{eq:F27c}) and (\ref{eq:F27e}), respectively. The details are omitted for brevity. This completes the proof of Theorem \ref{the:3}.
 \end{IEEEproof}

\textit{\underline{Remark} 3:} It is worth to noting from Theorem \ref{the:3} that the secrecy outage probability of both users decreases as the local computing bits increase, and reaches the minimum when ${\ell_k}=\ell_k^{\max }$. Hence, both users must have sufficient energy budget such that ${{E_k} > {{\left( {{\varsigma _k}c_k^3{{\left( {\ell_k^{\max }} \right)}^3}} \right)} \mathord{\left/ {\vphantom {{\left( {{\varsigma _k}c_k^3{{\left( {\ell_k^{\max }} \right)}^3}} \right)} {{T^2}}}} \right.
 \kern-\nulldelimiterspace} {{T^2}}}}, \forall k \in \left\{ {m,\;n} \right\}$ holds and the remaining ${L_k} - \ell_k^{\max }$ bits can be offloaded successfully. Moreover, for user $n$, the outage occurs constantly, i.e., ${{\rm{P}}_{so,n}}\rightarrow1$, when $p_m \rightarrow \infty$. Furthermore, due to the priority-based design, ${{\rm{P}}_{so,m}}$ is affected by $h_{AP,m}$ and $p_m$ while ${{\rm{P}}_{so,n}}$ is not only affected by $h_{AP,n}$ and $p_n$, but also is affected by $h_{AP,m}$ and $p_m$. The value of ${{\rm{P}}_{so,m}}$ is inversely proportional to $h_{AP,m}$ and $p_m$, while the value of ${{\rm{P}}_{so,n}}$ is inversely proportional to $h_{AP,n}$ and $p_n$ but is proportional to $h_{AP,m}$ and $p_m$.

\emph{\underline{Feasibility analysis}}: It is obvious that problem ($\mathbf{P2}$) is divided into two subproblems ($\mathbf{P2.1-m}$) and ($\mathbf{P2.1-n}$) due to the priority-based design. Thereby, the feasibility of problem ($\mathbf{P2}$) can be checked by solving the subproblems ($\mathbf{P2.1-m}$) and ($\mathbf{P2.1-n}$) whose solutions are provided in closed forms in Theorem \ref{the:3}. Based on this, it is easy to conclude that problem ($\mathbf{P2}$) is feasible if and only if ${E_k} - \frac{{{\varsigma _k}c_k^3{{\left( {l_k^{\max }} \right)}^3}}}{{{T^2}}} > 0,\;\forall k \in \left\{ {m,\;n} \right\}$ holds.

For summarizing, we present the details of our proposed optimal solution to problem ($\mathbf{P2}$) in Algorithm 2.

\begin{algorithm}[!t]    
\label{alg:Algorithm 2}     
\caption{Optimal solution to problem ($\mathbf{P2}$)}   
\begin{algorithmic}[1]
\STATE \textbf{Setting}\\ $B$, $T$, $\alpha$, $\alpha_m, \alpha_n$, $\varsigma_m, \varsigma_n$, $c_m, c_n$, $L_m, L_n$, $E_m, E_n$;\\
channel condition: $\gamma_{AP,m}, \gamma_{AP,n}$ and $\gamma_{e,k}$;
\STATE \textbf{Initialization} \\ ${\ell_m}$ and ${\ell_n}$;
\STATE calculate ${\bm{\ell}}$, ${\bf{p}}$, ${{\bf{R}}_t}$ and ${{\bf{R}}_s}$ through Theorem \ref{the:3};
\STATE \textbf{output} \\$\ell_{m}^*, \ell_{n}^*$, $p_{m}^*, p_{n}^*$, $R_{s,m}^*, R_{s,n}^*$, $R_{t,m}^*$ and $R_{t,n}^*$.
\end{algorithmic}
\end{algorithm}

\emph{\underline{Complexity analysis}}: It is easy to note that the complexity of Algorithm 2 is rather low, which mainly comes from the calculation of ${\bm{\ell}}$, ${\bf{p}}$, ${{\bf{R}}_t}$ and ${{\bf{R}}_s}$ in step 3. From the closed-form solution given in Theorem \ref{the:3}, similar to the complexity analysis for Algorithm 1, only 26 multiplications and 7 additions are required for Algorithm 2.

\section{Numerical Results}

In this section, numerical results are provided to validate the performance of our proposed design compared to two benchmark schemes, as well as one conventional design without an eavesdropper.

\textit{1) Secure full offloading:} All the users choose to offload all the task input bits to the AP in the considered secure NOMA-MEC system. This scheme corresponds to solve problem (P1) and (P2) by setting ${\ell_k} = 0,\;\forall k \in \left\{ {m,\;n} \right\}$.

\textit{2) Secure OMA-MEC offloading:} Two users adopt the TDMA protocol for computation offloading partially, where the time duration $T$ is divided into two parts with one part occupied by user $m$ and the rest by user $n$. The optimization of time slot allocation among two users is also taken into account.

\textit{3) Conventional design without eavesdropper:} No eavesdropper exists in our considered secure NOMA-MEC system, this corresponds to the scenario where ${h_{e,k}} = 0,\;\forall k \in \left\{ {m,\;n} \right\}$.

Here, we do not consider the \textit{local computing only scheme} \cite{IEEEhowto:11} since the computation tasks are locally performed instead of offloading to the MEC server for computing. In this case, it does not need to consider the secure issue.

The simulation parameters are set based on the works in \cite{IEEEhowto:4, IEEEhowto:20, IEEEhowto:21}. We set the system bandwidth for computation offloading as $B = 1$ MHz, the time duration as $T = 0.1$ sec, path-loss exponent as $\alpha  = 4$, the noise variance as $\sigma _{AP}^2 = \sigma _e^2 =  - 70$ dBm, the CPU cycles as ${c_m} = {c_n} = {10^3}$ cycles/bit, the effective capacitance coefficient as ${\varsigma _m} = {\varsigma _n} = {10^{ - 28}}$, the number of the computation input bits as ${L_k} = 2 \times {10^5}$ bits and $\ell_k^{\max } = 1.6 \times {10^5}$ bits, $\forall k \in \left\{ {m,\;n} \right\}$, the distance as ${d_{AP,m}} = {d_{AP,n}} = 60$ meters and ${d_{e,m}} = {d_{e,n}} = 100$ meters, and the secrecy outage probability as $\varepsilon  = 0.1$. Unless otherwise noted, the default parameters are given as mentioned above. The numerical results are obtained by averaging over 1000 random channel realizations.

\subsection{Weighted Sum-energy Consumption Minimization}

\begin{figure}[!t]
\centering
\includegraphics[width=3.3in]{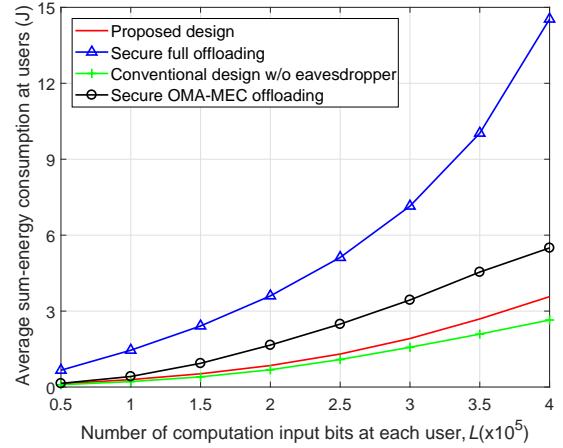}
\caption{Average sum-energy consumption at two users versus the number of computation input bits $L$ at each user.}
\label{fig:simulation1}
\end{figure}

Fig. \ref{fig:simulation1} shows the average sum-energy consumption of the two users versus the number of computation input bits ${L_k} = L,\;\forall k \in \left\{ {m,\;n} \right\}$ for each user, where $\ell_k^{\max }$ is set to be $0.8L$. It is observed that three partial offloading schemes (the proposed design, the secure OMA-MEC offloading and the conventional design w/o eavesdropper) achieve lower average sum-energy consumption than the full offloading scheme (the secure full offloading). This validates the benefit of the partial offloading mode by exploiting both the resources of local computation and the MEC server in the case where the given secrecy outage performance is satisfied. Such an advantage is further expanded when the number of computation input bits becomes large. Moreover, it can also be observed that by introducing the advanced NOMA technology our proposed design outperforms the secure the OMA-MEC offloading scheme. Nevertheless, as expected  it consumes more energy in our proposed design for the purpose of anti-eavesdropping than in the conventional design without an eavesdropper. In the small $L$ region (e.g., $L \le 1 \times {10^5}$), we observe that the proposed design, the conventional design without an eavesdropper and the secure OMA-MEC offloading achieve the similar energy consumption performance while they all outperform the secure full offloading scheme. This indicates the local computing is a more energy-efficient option in processing the computation tasks.

\begin{figure}[!t]
\centering
\includegraphics[width=3.3in]{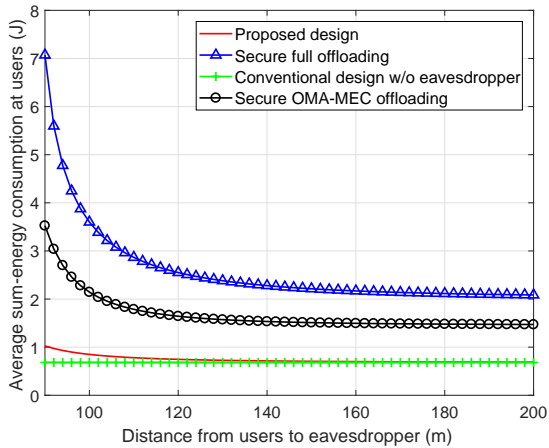}
\caption{Average sum-energy consumption at two users versus the distance ${d_e}$ from the users to the eavesdropper.}
\label{fig:simulation2}
\end{figure}

Fig. \ref{fig:simulation2} shows the average sum-energy consumption of the two users as a function of  the distance between the users and the eavesdropper, where ${d_{e,m}} = {d_{e,n}} = {d_e}$. It is observed that the consumed energy decreases as the distance increases due to the weakening of the wiretap channels, and gradually converges  to a constant value. More specifically, the proposed design shows a similar energy consumption performance as the conventional design without an eavesdropper when the distance is longer than $120$ m, and has a much better performance than the secure OMA-MEC offloading scheme and the secure full offloading scheme. We also observe that the gap of the energy consumption performance between the proposed design and the secure OMA-MEC offloading is relatively large when ${d_e} \le 110$ m, and the gap further increases as the distance decreases. This demonstrates the outstanding advantage of the NOMA assisted MEC system compared with the OMA counterpart in terms of anti-eavesdropping, especially in the strong eavesdropping case.

\begin{figure}[!t]
\centering
\includegraphics[width=3.3in]{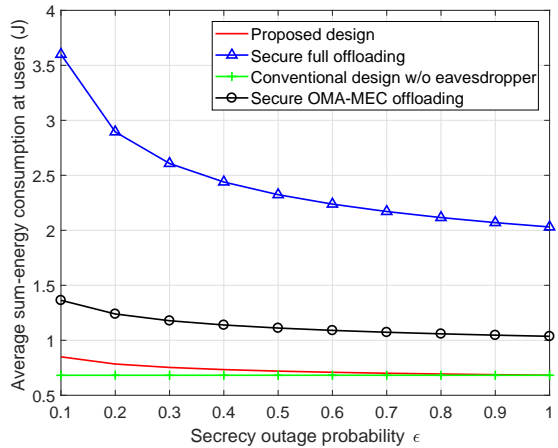}
\caption{Average sum-energy consumption at two users versus the secrecy outage probability of each user.}
\label{fig:simulation3}
\end{figure}

Fig. \ref{fig:simulation3} shows the average sum-energy consumption of the two users as a function of  the outage probability $\varepsilon $ of each user. In general, we have the similar observation as shown in Fig. \ref{fig:simulation2}. When the secrecy outage probability increases, the sum-energy consumption for secure offloading decreases, as the requirement for secrecy offloading becomes lower. And the proposed design is observed to have a similar performance as the conventional design without an eavesdropper when $\varepsilon $ is larger than $0.5$. This indicates that the introduced secrecy outage probability is a suitable metric to capture the secrecy offloading performance of our proposed NOMA-MEC system. In addition, we also observe that our proposed design is superior to both the secure OMA-MEC offloading and the secure full offloading schemes.

\subsection{Secrecy Outage Probability Minimization}

\begin{figure}[!t]
\centering
\includegraphics[width=3.3in]{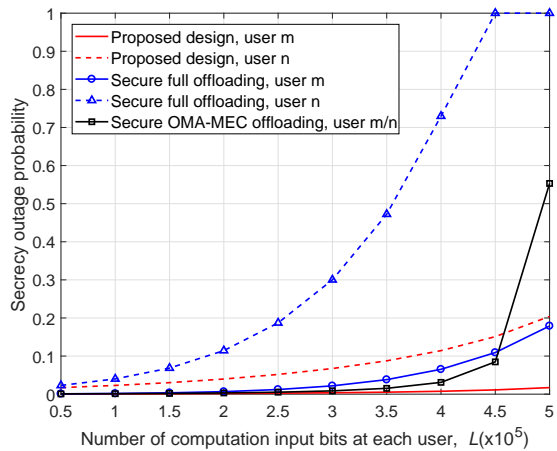}
\caption{Secrecy outage probability versus the number of computation input bits $L$ at each user.}
\label{fig:simulation4}
\end{figure}


Fig. \ref{fig:simulation4} shows the secrecy outage probability of each user versus the number of computation input bits $L$ for each user, where ${E_m} = {E_n} = E = 0.55$ Joule. For the secure OMA-MEC offloading, user $m$ and user $n$ share the same secrecy outage probability since the time duration $T$ is divided into two parts of the same size for each of them. In this figure, user $m$ shows better secrecy outage performance than user $n$. This indicates that the adopted priority based design works and user $m$ indeed enjoys higher priority than user $n$. It is observed that when $L \le 2 \times {10^5}$ bits, the proposed design, the secure full offloading and the secure OMA-MEC offloading achieve almost the same and extremely high secrecy outage performance for user $m$. However, when $L$ becomes large (e.g., $L \ge 2.5 \times {10^5}$ bits), the proposed design is observed to continue the good secrecy outage performance and outperform the other two schemes. The reason is that $0.55$ Joule energy budget for user $m$ is sufficient to support the secure full offloading or the secure partial offloading when the amount of computation tasks is not too heavy. However, as the amount of the computation tasks increases, the energy budget is incompetent in supporting high level secrecy offloading for neither NOMA based full offloading nor OMA based partial offloading. This verifies the importance of NOMA based partial offloading for information security. For user $n$, it is observed that the proposed design has better secrecy outage performance than the secure OMA-MEC offloading when $L > 4.5 \times {10^5}$ bits as well as the secure full offloading.

\begin{figure}[!t]
\centering
\includegraphics[width=3.3in]{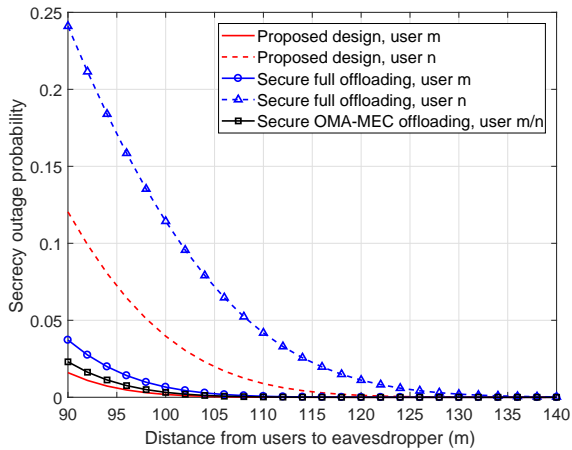}
\caption{Secrecy outage probability versus the distance ${d_e}$ from the users to the eavesdropper.}
\label{fig:simulation5}
\end{figure}

Fig. \ref{fig:simulation5} shows the secrecy outage probability of each user versus the identical distance from the users to the eavesdropper, where ${d_{e,m}} = {d_{e,n}} = {d_e}$ and $E = 0.5$ Joule. Similar to the observation in Fig. \ref{fig:simulation2}, the secrecy outage probability of three schemes decreases as the distance increases. From this figure, for user $m$, we observe that the proposed design is superior to the secure full offloading and the secure OMA-MEC offloading when distance ${d_e} \le 104$ m. Notably, the performance gap among them shrinks gradually as ${d_e}$ grows and eventually goes to zero. This verifies the critical influence of distance from the users to eavesdropper in secure offloading. Although the eavesdropper's instantaneous CSI is unknown at the users, the approximate perfect secrecy is achievable when distance ${d_e}$ becomes large. Such a characteristic is particularly distinct in our proposed design, even for user $n$.

\begin{figure}[!t]
\centering
\includegraphics[width=3.3in]{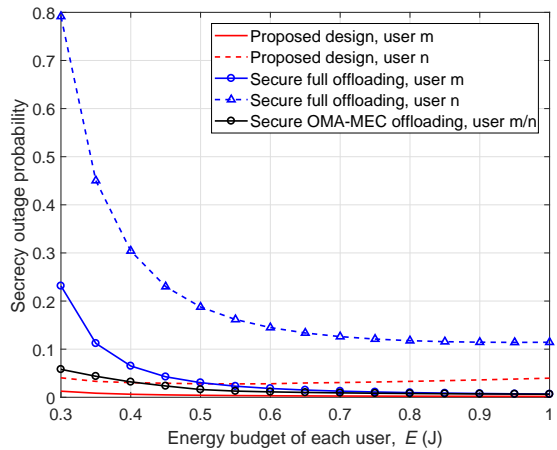}
\caption{Secrecy outage probability versus the energy budget $E$ of each user.}
\label{fig:simulation6}
\end{figure}

Fig. \ref{fig:simulation6} shows the secrecy outage probability of each user versus the energy budget $E$ of each user. As shown in this figure, when $E$ is small (e.g., $E < 0.45$ Joule), the proposed design is observed to outperform the secure OMA-MEC offloading for both user $m$ and user $n$. By contrast, when $E$ is further increased (e.g., $E > 0.45$ Joule), the user $n$'s secrecy outage performance of our proposed design degrades gradually and becomes worse than the secure OMA-MEC offloading. This is due to the fact that, as shown in (\ref{eq:F25}), user $m$'s transmission power affects user $n$'s secrecy outage probability adversely. For user $n$, such a negative effect  suffers in the low energy budget region and the secrecy outage performance gradually improves with its transmission power. But, the resulting interference exceeds the tolerance of user $n$ as user $m$'s transmission power further increases, which inevitably results in the degradation in secrecy outage performance of user $n$. However,  this recession is not significant due to the positive impact of user $n$'s transmission power. From this figure, we also realize that full offloading is not a good choice to ensure security.

\section{Conclusions}

In this paper, we studied an uplink NOMA-enabled MEC-aware network in the presence of a malicious eavesdropper, where two users simultaneously offload their partial computation tasks to the AP over the same resource block under secrecy considerations. We first obtained the optimal computing and communication resource allocations with semi-closed form expressions for two users to minimize the weighted sum-energy consumption. Then, we focused on the secrecy outage probability minimization problem by taking the priority of two users into account, and characterized the optimal secrecy offloading rates and power allocations with closed-form expressions. Numerical results validated the correctness of our theoretical analysis and demonstrated the advantages of our proposed designs over some other existing schemes.

\appendices
\section{Proof of Theorem \ref{the:1}}
We first focus on the secrecy outage probability ${{\rm{P}}_{so,n}}$. By substituting ${R_{t,n}}$ in (\ref{eq:F12}) into (\ref{eq:F10}), ${{\rm{P}}_{so,n}}$ can be re-expressed as
\begin{equation}\label{eq:F15}
\begin{array}{l}
{{\rm{P}}_{so,n}} \\
=\Pr\left\{ {{R_{t,n}} - {R_{s,n}} < {C_{e,n}}} \right\}\\
=\Pr\left\{ {{{\log }_2}\left( {1 + \frac{{{\gamma _{AP,n}}{p_n}}}{{1 + {\gamma _{AP,m}}{p_m}}}} \right) - {R_{s,n}} < {{\log }_2}\left( {1 + {\gamma _{e,n}}{p_n}} \right)} \right\}\\
= \Pr\left\{ {{{\left| {{h_{e,n}}} \right|}^2} > {\phi _n}} \right\}.
\end{array}
\end{equation}
where ${\phi _n} = \frac{{1 + {\gamma _{AP,m}}{p_m} + {\gamma _{AP,n}}{p_n} - \left( {1 + {\gamma _{AP,m}}{p_m}} \right){2^{{R_{s,n}}}}}}{{\left( {1 + {\gamma _{AP,m}}{p_m}} \right){2^{{R_{s,n}}}}{p_n}}}\sigma _e^2$. Recall that the probability density function (PDF) of ${\left| {{h_{e,n}}} \right|^2}$ is ${f_{{{\left| {{h_{e,n}}} \right|}^2}}}\left( x \right) = d_{e,n}^\alpha {e^{ - d_{e,n}^\alpha x}}$.
Then, ${{\rm{P}}_{so,n}}$ can be calculated as
\begin{equation}\label{eq:F16}
\begin{array}{l}
{{\rm{P}}_{so,n}} = \Pr\left\{ {{{| {{h_{e,n}}} |}^2} > {\phi _n}} \right\}\\
~\quad \;\;{\kern 1pt}{\kern 1pt}{\kern 1pt}  = \int_{{\phi _n}}^{ + \infty } {d_{e,n}^\alpha {e^{ - d_{e,n}^\alpha x}}} dx\\
~\quad \;\;{\kern 1pt}{\kern 1pt}{\kern 1pt}  = \left. { - {e^{ - d_{e,n}^\alpha x}}} \right|_{{\phi _n}}^{ + \infty }\\
~\quad \;\;{\kern 1pt}{\kern 1pt}{\kern 1pt}  = {e^{ - {\phi _n}d_{e,n}^\alpha }}.
\end{array}
\end{equation}
Hence, substituting (\ref{eq:F16}) into (\ref{eq:F11e}), we have ${e^{ - {\phi _n}d_{e,n}^\alpha }} \le \varepsilon $. After some basic mathematical transformations, the inequality in (\ref{eq:F14a}) is immediately obtained.

In the following, we will rewrite the secrecy outage probability ${{\rm{P}}_{so,m}}$. Similarly, substituting ${R_{t,m}}$ in (\ref{eq:F12}) into (\ref{eq:F10}) and note that the PDF of ${\left| {{h_{e,m}}} \right|^2}$ is ${f_{{{\left| {{h_{e,m}}} \right|}^2}}}\left( x \right) = d_{e,m}^\alpha {e^{ - d_{e,m}^\alpha x}}$, we have
\begin{equation}\label{eq:F17}
\begin{array}{l}
{{\rm{P}}_{so,m}} = \Pr\left\{ {{R_{t,m}} - {R_{s,m}} < {C_{e,m}}} \right\}\\
~\quad \;\;\;{\kern 1pt} {\kern 1pt}  = \Pr\left\{ {{{\left| {{h_{e,m}}} \right|}^2} > \frac{{1 + {\gamma _{AP,m}}{p_m} - {2^{{R_{s,m}}}}}}{{{2^{{R_{s,m}}}}{p_m}}}\sigma _e^2} \right\}\\
~\quad \;\;\;{\kern 1pt} {\kern 1pt}  = {\kern 1pt} {\kern 1pt} \int_{{\phi _m}}^{ + \infty } {d_{e,m}^\alpha {e^{ - d_{e,m}^\alpha x}}} dx\\
~\quad \;\;\;{\kern 1pt} {\kern 1pt}  = \left. { - {e^{ - d_{e,m}^\alpha x}}} \right|_{{\phi _m}}^{ + \infty }\\
~\quad \;\;\;{\kern 1pt} {\kern 1pt}  = {e^{ - {\phi _m}d_{e,m}^\alpha }},
\end{array}
\end{equation}
where ${\phi _m} = \frac{{1 + {\gamma _{AP,m}}{p_m} - {2^{{R_{s,m}}}}}}{{{2^{{R_{s,m}}}}{p_m}}}\sigma _e^2$. Then, according to the inequation in (\ref{eq:F11e}) and combined with the result given in (\ref{eq:F17}), the inequality in (\ref{eq:F14b}) can be derived straightly. This completes the proof of Theorem \ref{the:1}.



\ifCLASSOPTIONcaptionsoff
  \newpage
\fi

\end{document}